\documentclass[12pt]{article}
\usepackage{footnote}
\usepackage{empheq}

\catcode`\@=11
\@addtoreset{equation}{section}

\usepackage{amsmath}
\usepackage{amsbsy}
\usepackage{amsfonts}
\usepackage{latexsym}
\usepackage{mathrsfs}
\usepackage{graphicx}
\usepackage{marvosym}
\usepackage{url}

\newcommand{\beq}{\begin{equation}}
\newcommand{\eeq}{\end{equation}}
\newcommand{\bea}{\begin{eqnarray}}
\newcommand{\eea}{\end{eqnarray}}

\def\Tr{\mathop{\rm Tr}}

\newcommand{\be}{\begin{equation}}
\newcommand{\ee}{\end{equation}}

\newcommand{\de}{\partial}
\newcommand{\m}{\mu}
\newcommand{\n}{\nu}
\newcommand{\s}{\sigma}
\newcommand{\g}{\gamma}

\newcommand{\ve}{\epsilon}
\newcommand{\w}{\omega}
\newcommand{\1}{_{\text{\tiny{(1)}}}}
\newcommand{\2}{_{\text{\tiny{(2)}}}}

\newcommand{\chir}{\bar}

\newcommand{\Pp}{P_+}
\newcommand{\Pm}{P_-}
\newcommand{\Id}{1 \hspace{-2.7pt} \text{l}}
\newcommand{\Q}{\mathcal{Q}}
\newcommand{\N}{\mathcal{N}}
\newcommand{\li}{{(\ell)}}

\newcommand{\susy}{\mathbf{Q}}

\def\XXint#1#2#3{{\setbox0=\hbox{$#1{#2#3}{\int}$}
     \vcenter{\hbox{$#2#3$}}\kern-.5\wd0}}

\begin{document}
%
%
\begin{titlepage}

\begin{flushright}
\normalsize
~~~~
SISSA  59/2014/FISI-MATE
\end{flushright}

\vspace{80pt}

\begin{center}
{\LARGE $\N=2$ supersymmetric gauge theories on $S^2\times S^2$\\ and 
\vspace{0.4cm}
 \\ Liouville Gravity
}
\end{center}

\vspace{25pt}

\begin{center}
{
Aditya Bawane, Giulio Bonelli, Massimiliano Ronzani and Alessandro Tanzini 
}\\
%
\vspace{15pt}
%
International School of Advanced Studies (SISSA) \\via Bonomea 265, 34136 Trieste, Italy 
and INFN, Sezione di Trieste \footnote{email: abawane,bonelli,mronzani,tanzini@sissa.it}\\

\end{center}
%
\vspace{20pt}
%
We consider $\N=2$ supersymmetric gauge theories on four manifolds admitting an isometry. 
Generalized Killing spinor equations are derived from the consistency of supersymmetry
algebrae and solved in the case of four manifolds admitting a $U(1)$ isometry.
This is used to explicitly compute the supersymmetric path integral on $S^2\times S^2$ via equivariant localization.
The building blocks of the resulting partition function are shown to contain the three point functions and 
the conformal blocks of Liouville Gravity.



\vfill

\setcounter{footnote}{0}
\renewcommand{\thefootnote}{\arabic{footnote}}

\end{titlepage}

\tableofcontents

\section{Introduction}
\label{sec:intro}

Understanding non-perturbative corrections in quantum field theories is an important problem in modern theoretical physics.
Supersymmetric quantum field theories are a particular subset of quantum field theories where it is possible to study and exactly quantify these effects.
This is due to peculiar non-renormalization theorems applicable to some privileged sectors 
in the space of observables which are protected by supersymmetry.
Supersymmetric quantum field theories can be placed in curved background space-time manifolds and probed by gravitational couplings.
This can be done under the condition that some supercharge survives the gravitational coupling, that is if some spinorial parameters
exist on the whole manifold and satisfy the proper generalized Killing spinor 
equations \cite{Dumitrescu:2012ha,Dumitrescu:2012at,Hama:2012bg,Klare:2013dka,Gupta:2012cy}. 

The spinorial parameters,
on top of being sections of the spinor bundles, are also sections of the 
R-symmetry bundle. 
It is well known 
that in the case of $\N=2$ extended supersymmetry 
an appropriate choice of the R-symmetry
bundle exists which preserves a fraction of supersymmetry. This is called topological twist \cite{Witten:1988ze}. 
In this framework the supersymmetry protection is expressed in terms
of cohomological triviality, and path integrals are typically localized on a finite dimensional reduced phase space (the moduli space of protected 
vacua configurations) via localization.

In the case of supersymmetric theories in two and four dimensions, this choice of the R-symmetry bundle revealed the supersymmetric theory
to be also an important instrument for predicting and obtaining very interesting solutions to counting problems in algebraic and differential geometry ,
e.g. Gromov-Witten \cite{Witten:1991zz} and Donaldson invariants \cite{Witten:1988ze}.

In order to exploit the full power of supersymmetry and reach explicit expressions of the quantities under study, it is important to solve
the moduli space integrals. This can be technically very difficult or impossible in the most general cases.
However, if the space-time manifold admits some isometry, one might be able to further localize the path integral over an invariant 
locus of the moduli space, thus improving upon the method of equivariant localization. For example, this allowed the exact computation of the instanton
partition function of ${\mathcal N}=2$ supersymmetric theories on $\mathbb{R}^4$ \cite{Nekrasov:2002qd,Bruzzo:2002xf}. The extension of this program to
toric compact manifolds  has been sketched in \cite{Nekrasov:2003vi}, where it was also conjectured that the resulting partition function
and observables provide a contour integral representation for Donaldson invariants.  More recently, $\N=2$ supersymmetric gauge theories
on the minimal resolution general toric singularities have been considered \cite{Bonelli:2012ny} in the context of 
AGT correspondence with two-dimensional
(super-)conformal field theories \cite{Alday:2009aq}. A rigorous mathematical framework for $A_n$ singularities has been presented
in \cite{Bruzzo:2013daa,Bruzzo:2014jza}.
Supersymmetric gauge theories with $\N=2$ on curved four manifolds have been considered for example in
\cite{Samtleben:2012gy,
Razamat:2013jxa,
Bobev:2013cja,
Closset:2013sxa,
Assel:2014paa,
Gomis:2014woa,
Marmiroli:2014ssa}. 
The specific case of $S^2\times S^2$ will also be considered in \cite{GM}.

In this paper we study $\N=2$ gauge theories on arbitrary Riemannian four manifolds, and show that supersymmetric parameters
satisfy generalized Killing spinor equations arising from the requirement of closure of the superalgebra.
For manifolds admitting an isometry, we show that these equations are solved by an
equivariant version of the topological twist
and we explicitly compute the gauge theory path-integral, which turns out to be given by an appropriate gluing of Nekrasov partition functions.

An interesting byproduct of our analysis is the natural appearance, in the $U(2)$ case, of three-point numbers and conformal blocks of Liouville gravity 
as building blocks of the $S^2\times S^2$ partition function, related respectively to the one-loop and the instanton sectors.
As we will discuss in Sect.5, a first hint to the relation with Liouville gravity can be obtained by considering the compactification
of two M5-branes on $S^2\times S^2\times\Sigma$. The central charge
of the resulting two-dimensional conformal field theory on $\Sigma$ can be computed from the M5-branes anomaly polynomial \cite{Bonelli:2009zp,Alday:2009qq}  
and is indeed consistent with our findings.   

Let us underline that our method applies in general to four-manifolds admitting a $U(1)$-action generated by a vector field $V$.
The path integral localizes on flat connections when $V$ has no zeros, for example Hopf surfaces or $S^1\times M_3$, otherwise
it localizes on (anti-)instantons on the zeros of $V$, as $S^4$ or compact toric manifolds. The case $S^2\times S^2$ 
discussed in detail in this paper belongs to the latter class.

The paper is organized as follows.
In Section 2 we discuss supersymmetry on curved four manifolds and derive the generalized Killing spinor equations from the superalgebra.
In Section 3 we obtain some relevant solutions of these equations on $S^2\times S^2$.
In Section 4 we use the results of the previous Sections to compute the partition function of the supersymmetric gauge theory on $S^2\times S^2$.
In Section 5 we compare the gauge theory computations with Liouville Gravity.
In Section 6 we discuss our results and comment on further developments.
Appendix A contains the detailed derivation of the full $\N=2$ supersymmetry generalized Killing equations discussed in Section 2.
Appendix B describes the solutions to Killing spinor equations in the general case of a four-manifold admitting a
$U(1)$ isometry.
Appendix C describes a set of other solutions to the latter that we report, but do not use in the main construction.
Appendix D contains our conventions on metric and spinors.
Appendix E contains our conventions on special functions.

\section{Supersymmetry on curved space}
\label{susy}

The algebras for  $\mathcal{N} = 1$ and  $\mathcal{N} = 2$ supersymmetry on four dimensional curved spaces have been recently derived using
supergravity considerations \cite{Dumitrescu:2012ha,Dumitrescu:2012at,Hama:2012bg,Klare:2013dka}.
In this section, we intend to re-derive the same results
in a direct way building on the consistency of the supersymmetry algebra.
For completeness and illustration of the method, we start by considering chiral ${\cal N}=1$ supersymmetry
and then we move to the full ${\cal N}=2$ supersymmetry algebra.

\subsection{${\cal N}=1$ Supersymmetry
}
We consider the case of supersymmetry algebra with one supercharge, parametrized by a 
(commuting) chiral spinor $\xi_\alpha$ of R-charge $+1$, and derive the algebra as realized on a vector multiplet, 
consisting of a gauge field $A_{\alpha\dot\alpha}$, gauginos $\lambda_\alpha$ and $\tilde\lambda_{\dot\alpha}$, and an auxiliary field $D$.

Supersymmetric variation of the gauge field and the gauginos is fixed by Lorentz covariance and R-charge conservation to be:
\begin{equation}
\begin{aligned}
&\delta A_{\alpha\dot\alpha} = \xi_\alpha\tilde\lambda_{\dot\alpha}, \\
&\delta\tilde\lambda_{\dot\alpha} =0, \\
&\delta\lambda_\alpha = i\xi_\alpha D +(F^+)_{\alpha\beta}\xi_\beta.
\label{eq:VectorOneSupercharge}
\end{aligned}
\end{equation}
Considering now the square of the supersymmetric variation of $\lambda_\alpha$, we get
\begin{equation}
\begin{aligned}
\delta^2 \lambda_\alpha &= i\xi_\alpha\delta D + [D_A(\xi\tilde\lambda)]\xi_\beta \\
&= i\xi_\alpha\delta D + \nabla_{(\alpha\dot\gamma}\xi_{\beta)}\tilde\lambda_{\dot\gamma}\xi_{\beta} + (\xi_{(\alpha}D_{\beta)\dot\gamma}\tilde\lambda_{\dot\gamma})\xi_\beta \\
&= i\xi_\alpha\delta D + \nabla_{(\alpha\dot\gamma}\xi_{\beta)}\tilde\lambda_{\dot\gamma}\xi_{\beta} + (\xi_{\alpha}D_{\beta\dot\gamma}\tilde\lambda_{\dot\gamma})\xi_\beta 
\end{aligned}
\end{equation}
where $\xi^2 = 0$ has been used to obtain the final term, $\nabla$ is the covariant derivative containing the spin connection and $D=\nabla+A$. 
We now notice that for the final expression to vanish, the middle term should align in the direction of $\xi_\alpha$,
so that all the terms can be compensated by $\delta D$. For this to happen, we are forced to require that
\begin{equation}
\nabla_{(\alpha\dot\gamma}\xi_{\beta)} = i\hat V_{\alpha\dot\gamma}\xi_\beta + i\hat V_{\beta\dot\gamma}\xi_\alpha
\end{equation}
for some background connection $\hat V$. We note that this is equivalent to the Killing spinor equation
\begin{equation}
\nabla_{\alpha\dot\alpha} \xi_\beta = i V_{\alpha\dot\alpha}\xi_\beta + i W_{\beta\dot\alpha}\xi_\alpha,
\label{n=1}
\end{equation}
where $\hat V=V+W$.
Requiring this allows us to set $\delta^2 \lambda_\alpha = 0$ if we set
\begin{equation}
\delta D = iD_{\beta\dot\gamma}\tilde\lambda_{\dot\gamma}\xi_\beta - \hat V_{\beta\dot\gamma}\tilde\lambda_{\dot\gamma}\xi_\beta.
\end{equation}
It follows from a routine calculation that $\delta^2 D=0$.

Notice that (\ref{n=1}) is the Killing spinor equation derived in \cite{Klare:2012gn,Dumitrescu:2012ha}.

The same equation can be derived by considering the chiral multiplet in the following way.
The supersymmetry variations of an anti-chiral multiplet $(\tilde\phi, \tilde\psi_{\dot\alpha}, F)$ generated by one supercharge of R-charge $+1$
are
\begin{equation}
\begin{aligned}
&\delta\tilde\phi =0, \\
&\delta\tilde\psi_{\dot\alpha} = i\xi_\alpha D_{\alpha\dot\alpha}\tilde\phi, \\
&\delta \tilde F = i\xi_\alpha D_{\alpha\dot\alpha}\tilde\psi_{\dot\alpha} + \xi_\alpha[\lambda_\alpha,\tilde\phi] + \xi_\alpha V_{\alpha\dot\alpha}\tilde\psi_{\dot\alpha}
\end{aligned}
\end{equation}
Consider first the square of the variation of $\tilde\psi_{\dot\alpha}$:
\begin{equation}
\begin{aligned}
\delta^2 \tilde\psi_{\dot\alpha} &= i\xi_\alpha\left(  D_{\alpha\dot\alpha}\delta\tilde\phi  + [\delta A_{\alpha\dot\alpha},\tilde \phi] \right) \\
&= i\xi_\alpha\left(  D_{\alpha\dot\alpha}\delta\tilde\phi  + [\xi_\alpha\tilde\lambda_{\dot\alpha},\tilde \phi] \right) \\
&= 0
\end{aligned}
\end{equation}
since $\delta\tilde\phi =0$ and $\xi^2 = 0$. Consider similarly $\delta^2 \tilde F$:
\begin{equation}
\begin{aligned}
\delta^2 \tilde F 
=& i\xi_\alpha[\delta A_{\alpha\dot\alpha},\tilde\psi_{\dot\alpha}]
 + i\xi_\alpha D_{\alpha\dot\alpha}(i\xi_\beta D_{\beta\dot\alpha}\tilde\phi ) \\
&+ \xi_\alpha [ i\xi_\alpha D+(F^+)_{\alpha\beta}\xi_\beta, \tilde\phi]
 + \xi_\alpha V_{\alpha\dot\alpha}(i\xi_\beta D_{\beta\dot\alpha}\tilde\phi) \\
=&-\xi_\alpha\nabla_{\alpha\dot\alpha}\xi_\beta  D_{\beta\dot\alpha}\tilde\phi
 - \xi_\alpha\xi_\beta D_{\alpha\dot\alpha}D_{\beta\dot\alpha} \tilde\phi \\
&+ \xi_\alpha\xi_\beta [(F^+)_{\alpha\beta},\tilde\phi]
 + i\xi_\alpha V_{\alpha\dot\alpha}\xi_\beta D_{\beta\dot\alpha}\tilde\phi \\
=& -\xi_\alpha\nabla_{\alpha\dot\alpha}\xi_\beta  D_{\beta\dot\alpha}\tilde\phi
 + i\xi_\alpha V_{\alpha\dot\alpha}\xi_\beta D_{\beta\dot\alpha}\tilde\phi.
\end{aligned}
\end{equation}
This is vanishing by equation (\ref{n=1}).

\subsection{$\mathcal{N} = 2$ Supersymmetry}

We first consider the simpler case of chiral $\mathcal{N} =2$ supersymmetry.
Its straightforward (but tedious) generalization to the case with generators of both chiralities is treated next.

\subsubsection{Chiral $\mathcal{N} = 2$ Supersymmetry}

In this subsection, we derive the chiral $\mathcal{N}=2$ algebra generated by a doublet of
left-chirality spinors and the consistency conditions that the four manifold has to satisfy. 
We realize it on a vector multiplet. 
The derivation is based on the following considerations:
\begin{itemize}
 \item The supersymmetry transformations of the scalar fields
$\phi$,$\bar{\phi}$ and the vector field $A_\m$ 
are
\begin{eqnarray}
 \susy_L A_\m &=& i\xi^A\sigma_\m\bar\lambda_A,
 \nonumber \\
 \susy_L\phi &=& -i\xi^A\lambda_A,
 \nonumber \\
 \susy_L\bar\phi &=& 0.
\label{chiralvec}
\end{eqnarray}
\item The chiral supersymmetry transformation squares to a gauge transformation on
the vector multiplet. This implies the differential equations 
(``Killing spinor equations'') satisfied by the transformation parameter
$\xi_A$ in order to the supersymmetry to hold. 
The specific 
form of the generator of the
gauge transformation
will be derived in the following.
\item The scaling dimension of any background field is positive. The reason for
this assumption is that we would like to recover the familiar algebra in the
flat-space limit. 
Positivity of the scaling dimensions of background fields ensures that as the
characteristic length scales of the manifold go to infinity (or equivalently, as
we approach the flat metric),
the background fields go to zero.
\end{itemize}

We recall below the $U(1)_R$ charges and scaling dimensions of the fields
\begin{center}
\begin{tabular}{c c c c c c c c}
   Field & $\phi$ & $\bar\phi$ & $\lambda_A$ & $\bar\lambda_A$ & $A_\m$ & $D_{AB}$ & $\xi_A$\\ \hline
   $U(1)_R$ charge & $2$ & $-2$ & $1$ & $-1$ & $0$ & $0$ & $1$\\
\end{tabular}
 \end{center}
\begin{center}
\begin{tabular}{c c c c c c c}
   Field & $\phi$ & $\bar\phi$ & $\lambda_A$ & $\bar\lambda_A$ & $D_{AB}$ & $\xi_A$\\
\hline
   Scaling dimension & $1$ & $1$ & $3/2$ & $3/2$  & $2$ & $-1/2$\\
\end{tabular}
 \end{center}

As in the previous section, we now show how the closure of the supersymmetry algebra implies generalized Killing spinor equations
with background fields. 
The most general variation of $\bar\lambda_A$ consistent with the considerations above is
\begin{equation}
 \susy_L\bar\lambda_A =
  a\bar\sigma^\m\xi_A D_\m\bar\phi
 +b\bar\sigma^\m D_\m\xi_A\bar\phi
\end{equation}
where $a$ and $b$ are complex numbers to be determined.
Squaring supersymmetry, we get
\begin{equation}
  \susy_L^2\bar\lambda_A = a
\bar\sigma^\m\xi_A[\xi^B\sigma_\m\bar\lambda_B,\bar\phi] =
i[i\xi^B\xi_B\bar\phi,\bar\lambda_A] =:i [\Phi, \bar\lambda_A]
\end{equation}
where the last equality defines the generator of gauge transformations $\Phi =
i\xi^B\xi_B\bar\phi$. Consider now the square of the supersymmetry variation of
the gauge field
$A_\m$
\begin{equation}
  \susy_L^2A_\m = i\xi^A\sigma_\m\susy_L\bar\lambda_A = ia\xi^B\xi_BD_\m\bar\phi
+ ib\bar\phi \xi^B \sigma_\m \bar \sigma^\n D_\n\xi_B.
\end{equation}
Since the supersymmetry squares to gauge transformation, and since the generator
of gauge transformation is $\Phi = i\xi^B\xi_B\bar\phi$, we require that
\begin{equation}
 ia\xi^B\xi_BD_\m\bar\phi + ib\bar\phi \xi^B \sigma_\m \bar \sigma^\n D_\n\xi_B
= \susy_L^2A_\m = D_\m(i a\xi^B\xi_B\bar\phi)
\end{equation}
which gives
\begin{equation}
 2a\xi^B D_\m\xi_B = b \xi^B \sigma_\m \bar \sigma^\n D_\n\xi_B.
\end{equation}
We note that the above equation is satisfied when $a=2b$ and $D_\m\xi_A = \frac{1}{4}
\sigma_\m \sigma^\n D_\n\xi_A$ (or 
equivalently $D_\m\xi_A = \sigma_\m {\bar\xi}'_A$ for some right chirality
spinor ${\bar\xi}'_A$). To see that this is indeed the conformal Killing
equation, we consider the supersymmetry variation of $\lambda_A$.
The most
general expression possible is
\begin{equation}
\susy_L\lambda_A = \frac{1}{2}\sigma^{\m\n}\xi_A(kF_{\m\n}+\bar\phi T_{\m\n} +
\phi W_{\m\n})
  +c\,\xi_A[\phi,\bar\phi]
 +D_{AB}\xi^B
\label{eq:chiralgaugino}
\end{equation}
where $k$ and $c$ are complex numbers yet to be determined; $T_{\m\n}$ and
$W_{\m\n}$ are anti self-dual background fields, both having mass dimension $1$
and with $U(1)_R$ charge $2$ and $-2$ respectively.
Computing $\susy_L^2\phi$, we
immediately see that $c = ia$. After some algebra, we find
\begin{equation}
\begin{aligned}
\susy_L^2\lambda_A
&= i[ia\xi^B\xi_B\bar\phi,\lambda_A] + 2ik(\bar\lambda^B
  \bar\sigma^\m\xi_A)\left(D_\m\xi_B -\frac{1}{4}\sigma_\m\bar\sigma^\n D_\n\xi_B\right)
 -\frac{i}{2}\sigma^{\m\n}W_{\m\n}(\xi^B\lambda_B)\xi_A \\
&+\Big[\susy_LD_{AB}-ik(\xi_A\sigma^\m D_\m \bar\lambda_B + \xi_B\sigma^\m D_\m
\bar\lambda_A) - a[\bar\phi,\xi_A\lambda_B + \xi_B\lambda_A]\Big]\xi^B.
\label{eq:chiralsquaregaugino}
\end{aligned}
\end{equation}
The right hand side has been arranged in a form that allows some immediate
inferences. Firstly, the Killing spinor equation, as suggested earlier, is given
by
\begin{equation}
D_\m\xi_A = \sigma_\m {\bar\xi}'_A,
\label{chiraln=2}
\end{equation}
which also confirms that $a = 2b$. Noting that $a = 2b = ic$ can be absorbed
into $\bar\phi$, we will set $b=1$. Secondly, the background field $W_{\m\n}$
has to be zero since it cannot be absorbed into the variation of the auxiliary
field $D_{AB}$, which is symmetric in its indices and can not contain any term proportional to $\epsilon_{AB}$. 
Thirdly, we can read off the expression for $\susy_LD_{AB}$ by equating the last
parenthesis to zero:
\begin{equation}
\susy_LD_{AB} = ik(\xi_A\sigma^\m D_\m \bar\lambda_B + \xi_B\sigma^\m D_\m
\bar\lambda_A) + 2[\bar\phi,\xi_A\lambda_B + \xi_B\lambda_A].
\label{eq:chiralauxiliary}
\end{equation}
Finally, we look at the square of the chiral supersymmetry transformation of the
auxiliary field:
\begin{equation}
\begin{aligned}
\susy_L^2 D_{AB}
&= i[2i\xi^C\xi_C\bar\phi, D_{AB}] +
   4ik D^\m \bar\phi \left\{\xi_A \left(D_\m\xi_B -\frac{1}{4}\sigma_\m \bar\sigma^\n
   D_\n\xi_B \right) + (A \leftrightarrow B)\right\} \\
&+ ik\bar\phi\big\{\xi_A\sigma^\m \bar\sigma^\n D_\m D_\n\xi_B
 + (A\leftrightarrow B) \big\}=i[\Phi,D_{AB}].
\label{eq:squarechiralauxiliary}
\end{aligned}
\end{equation}
We recognize the first term to be the gauge transformation. The middle term in
the curly brackets is once again a contraction of the main equation
(\ref{chiraln=2}).
The last piece in curly brackets is new: its
vanishing is the additional condition on the Killing spinor
\begin{equation}
\xi_{(A}\sigma^\m\bar\sigma^\n D_\m D_\n \xi_{B)} = 0
\label{eq:auxKillingeqn}
\end{equation}
which implies 
\be\label{chiaux}
\sigma^\m\bar\sigma^\n D_\m D_\n \xi_{A} = M\xi_A
\ee 
for some scalar background field $M$. We call \eqref{chiaux}
the auxiliary equation. The leftover parameter $k$ can be set to one
by a rescaling of $\lambda_A$, $\phi$ and $D_{AB}$.
To summarize, the chiral supersymmetry transformation generated by a left
chirality spinor $\xi_A$ is given by
\be
\begin{aligned}
\susy_L A_\m &= i\xi^A\sigma_\m\bar\lambda_A,   \\
\susy_L\phi &= -i\xi^A\lambda_A,    \\
\susy_L\bar\phi &= 0,   \\
\susy_L\lambda_A &= \frac{1}{2}\sigma^{\m\n}\xi_A(F_{\m\n}+\bar\phi T_{\m\n})
                     +2i\xi_A[\phi,\bar\phi] +D_{AB}\xi^B, \\
\susy_L\bar\lambda_A &= 2\bar\sigma^\m\xi_AD_\m\bar\phi +\bar\sigma^\m D_\m\xi_A\bar\phi,  \\
\susy_LD_{AB} &= i(\xi_A\sigma^\m D_\m \bar\lambda_B + \xi_B\sigma^\m D_\m
                   \bar\lambda_A) + 2[\bar\phi,\xi_A\lambda_B + \xi_B\lambda_A]
\end{aligned}
\ee
where $\xi_A$ satisfies
\be
D_\m\xi_A - \frac{1}{4} \sigma_\m \bar\sigma^\n D_\n\xi_A = 0
\ee
\be
\sigma^\m\bar\sigma^\n D_\m D_\n \xi_{A} = M\xi_A
\ee
and $M$ is a scalar background field.

\subsubsection{Full $\mathcal{N}=2$ Supersymmetry}
\label{full}

We now turn to the case of $\mathcal{N}=2$ supersymmetry with generators of both chiralities.

In this case we start from 
the supersymmetry transformations of the scalar fields
and of the vector field
\be
\begin{aligned}
 \susy A_\m &= i\xi^A\sigma_\m\bar\lambda_A-i\bar\xi^A\bar\sigma_\m\lambda_A,
 \\
 \susy\phi &= -i\xi^A\lambda_A,
 \\
 \susy\bar\phi &= +i\bar\xi^A\bar\lambda_A,
\end{aligned}
\ee
where the right chirality spinor $\bar\xi^A$ has $U(1)_R$ charge $-1$ and mass dimension $-1/2$.

Moreover we exploit the fact that
the superconformal transformation squares to a sum of 
gauge transformation, Lorentz transformation, scaling, $U(1)_R$ transformation and $SU(2)_R$ transformation, generated by functions denoted by $\Phi$, $V$, $w$, $\Theta$ and 
$\Theta_{AB}$ respectively, whose expressions we will determine in the following. Notice that the generator of scaling transformations $w$ is related in four dimensions 
to the generator of coordinate translations $V$ as $w = \frac{1}{4}D_\m V^\m$.

Using the masses and charges tabulated earlier, we can write the form of $\susy^2$ for all members of the vector multiplet. For example
\begin{equation}
 \susy^2\lambda_A =
 \left(iV^\n D_\n \lambda_A +\frac{i}{4}(\sigma^{\m\n} D_\m V_\n)\lambda_A\right)+i[\Phi,\lambda_A]+\frac{3}{2}w\lambda_A +\Theta\lambda_A
 +\Theta_{AB}\lambda^B
\end{equation}
and for the gauge field
\begin{equation}
 \susy^2A_\m = i V^\n F_{\n\m}+D_\m\Phi
\end{equation}
and so on and so forth for the other members of the multiplet. 

The explicit computations are reported in Appendix A and the final results are as follows.
The spinor parameters have to satisfy the generalized Killing equation
\be
 D_\m \xi_B + T^{\rho\sigma}\sigma_{\rho\sigma}\sigma_\m\bar\xi_B - \frac{1}{4} \sigma_{\m}\bar\sigma_{\n}D^{\n}\xi_B=0
\ee
and 
\be
 D_\m \bar\xi_B + \bar T^{\rho\sigma}\bar\sigma_{\rho\sigma}\bar\sigma_\m\xi_B - \frac{1}{4} \bar\sigma_{\m}\sigma_{\n}D^{\n}\bar\xi_B=0
\ee
and the auxiliary equations
\begin{equation}
\begin{aligned}
\s^\m \chir\s^\n D_\m D_\n \xi_A + 4 D_\lambda T_{\m\n} \s^{\m\n} \s^\lambda \chir\xi_A &= M_1\xi_A, \\
\chir \s^\m \s^\n D_\m D_\n \chir\xi_A + 4 D_\lambda \chir T_{\m\n} \chir\s^{\m\n} \chir\s^\lambda \xi_A &= M_2\chir\xi_A.
\end{aligned}
\end{equation}

We summarize the supersymmetry algebra just derived for the vector multiplet
\be\label{Q-alg}
\begin{aligned}
\susy A_\m &=i\xi^A\s_\m \chir\lambda_A - i \chir\xi^A \chir\s_\m \lambda_A, \\
\susy\phi &=-i\xi^A\lambda_A, \\
\susy\chir\phi &=+i\chir\xi^A\chir\lambda_A, \\
\susy\lambda_A &=\frac{1}{2}\s^{\m\n}\xi_A(F_{\m\n}+8\chir\phi T_{\m\n})+2\s^\m\chir\xi_A D_\m\phi+\sigma^\m D_\m\bar\xi_A\phi+2i\xi_A[\phi,\chir\phi]+D_{AB}\xi^B, \\
\susy\chir\lambda_A &=\frac{1}{2}\chir\s^{\m\n}\chir\xi_A(F_{\m\n}+8\phi \chir T_{\m\n})+2\chir\s^\m \xi_A D_\m\chir\phi+\bar\sigma^\m D_\m\xi_A\bar\phi-2i\chir\xi_A[\phi,\chir\phi]+D_{AB}\chir\xi^B, \\
\susy D_{AB} &=
 -i\bar\xi_A\bar\sigma^mD_m\lambda_B
 -i\bar\xi_B\bar\sigma^mD_m\lambda_A
 +i\xi_A\sigma^mD_m\bar\lambda_B
 +i\xi_B\sigma^mD_m\bar\lambda_A
 \\ 
 &-2[\phi,\bar\xi_A\bar\lambda_B+\bar\xi_B\bar\lambda_A]
 +2[\bar\phi,\xi_A\lambda_B+\xi_B\lambda_A].
\end{aligned}
\ee
The square of the supersymmetry action is
\be\label{Qsq-alg}
\begin{aligned}
\susy^2 A_\m &= i\iota_V F+D \Phi, \\
\susy^2 \phi &= i\iota_V D\phi +i[\Phi,\phi]+(w+2\Theta)\phi, \\
\susy^2 \chir\phi &=i \iota_V D\chir\phi +i[\Phi,\chir\phi]+(w-2\Theta)\chir\phi, \\
\susy^2 \lambda_A &=i \iota_V D\lambda_A+i[\Phi,\lambda_A]+(\frac{3}{2}w+\Theta)\lambda_A+
     \frac{i}{4} (D_\rho V_\tau)\s^{\rho\tau}\lambda_A + \Theta_{AB}\lambda^B, \\
\susy^2 \chir\lambda_A &=i \iota_V D\chir\lambda_A+i[\Phi,\chir\lambda_A]+(\frac{3}{2}w-\Theta)\chir\lambda_A+
     \frac{i}{4}(D_\rho V_\tau)\chir\s^{\rho\tau}\chir\lambda_A  + \Theta_{AB}\chir\lambda^B, \\
\susy^2 D_{AB} &=i \iota_V D (D_{AB}) +i[\Phi,D_{AB}]+2wD_{AB}+\Theta_{AC}D^C_{\phantom{C}B}+\Theta_{BC}D^C_{\phantom{C}A},
\end{aligned}
\ee
where the parameters of the bosonic transformations are
\be\label{bosonic-par}
\begin{aligned}
V^\mu&=2\chir\xi^A\chir\s^\mu\xi_A, \\
\Phi&=2i\chir\xi_A\chir\xi^A\phi+2i\xi^A\xi_A\chir\phi, \\
\Theta_{AB}&=-i\xi_{(A}\s^\m D_\m\chir\xi_{B)}+iD_\m\xi_{(A}\s^\m\chir\xi_{B)}, \\
w&=-\frac{i}{2}(\xi^A\s^\m D_\m\chir\xi_A+ D_\m\xi^A\s^\m\chir\xi_A), \\
\Theta&=-\frac{i}{4}(\xi^A\s^\m D_\m\chir\xi_A- D_\m\xi^A\s^\m\chir\xi_A).
\end{aligned}
\ee

In general the spinorial parameters are sections of the corresponding
vector bundles, namely
$\xi\in\Gamma\left(S^+\otimes {\cal R}\otimes {\cal L}_R\right)$
and
$\bar\xi\in\Gamma\left(S^-\otimes {\cal R}^\dagger\otimes {\cal
L}_R^{-1}\right)$
where $S^\pm$ are the spinor bundles of chirality $\pm$,
${\cal R}$ is the $SU(2)$ R symmetry vector bundle
and
${\cal L}_R$ is the $U(1)$ R symmetry line bundle.
The four manifold is subject to the condition that the above product
bundles are well defined\footnote{Also the auxiliary
field $D\in\Gamma_S\left(\mathcal{R}\otimes \mathcal{R}\right)$ has to be well defined.}
and that a solution to the generalized Killing spinor equations
exists and is everywhere well defined.
These conditions differently constrain the space-time four manifold
depending on the choice of ${\cal R}$ and ${\cal L}_R$.
The choice leading to the topologically twisted theory
is to set ${\cal L}_R=\mathcal{O}$ to be the trivial line bundle and ${\cal R}=S^-$.
Therefore, for this choice of the R-symmetry bundles,
$S^+\otimes S^-\sim T$ and $S^-\times S^- \sim \mathcal{O}+ T^{(2,+)}$
with $T$ the tangent bundle and $T^{(2,+)}$ the bundle of selfdual forms.
In this case the four manifold has to be Riemannian and with a
Killing vector in order to admit this realization of the $\N=2$ super-algebra.

\section{Spinor solutions on $S^2\times S^2$}
\label{spinor}

As derived in the previous section, the conformal Killing spinors satisfy two sets of equations: the main equations
\be
\begin{aligned}\label{eq:1}
D_\m \xi_A &= - T^{\kappa\lambda}\s_{\kappa\lambda}\s_\m \chir{\xi}_A  -i \s_\m\chir{\xi}'_A, \\
D_\m \chir\xi_A &= - \chir T^{\kappa\lambda}\chir\s_{\kappa\lambda}\chir\s_\m \xi_A  -i \chir\s_\m \xi'_A, 
\end{aligned}
\ee
and the auxiliary equations\footnote{Here and in the following we consider the particular case $M_1=M_2=M$. 
This choice reproduces the auxiliary equations considered in \cite{Hama:2012bg}.}
\be
\begin{aligned}\label{eq:2}
\s^\m \chir\s^\n D_\m D_\n \xi_A + 4 D_\lambda T_{\m\n} \s^{\m\n} \s^\lambda \chir\xi_A &= M\xi_A \\
\chir \s^\m \s^\n D_\m D_\n \chir\xi_A + 4 D_\lambda \chir T_{\m\n} \chir\s^{\m\n} \chir\s^\lambda \xi_A &= M\chir\xi_A,
\end{aligned}
\ee
with
\be\label{eq:Dcov}
\begin{aligned}
D_\m \xi_A &=\nabla_\m\xi_A+ i {G_\m}_A^{\;\;B} \xi_B=\de_\m \xi_A + \frac{1}{4}\omega_\m^{ab}\s_{ab} \xi_A
+ i {G_\m}_A^{\;\;B} \xi_B, \\
D_\m \chir\xi_A &=\nabla_\m\chir\xi_A+ i {G_\m}_A^{\;\;B} \chir\xi_B= \de_\m \chir\xi_A + \frac{1}{4}\omega_\m^{ab}\chir\s_{ab} \chir\xi_A
+ i {G_\m}_A^{\;\;B} \chir\xi_B,
\end{aligned}
\ee
where $T^{\kappa\lambda}$ and $\chir T^{\kappa\lambda}$ are anti self-dual and self-dual background fields respectively,
$M$ is a scalar background field and $A,B,\dots$ are the $SU(2)_R$ doublet indices.
The covariant derivative involves spin connection $\omega_\m^{ab}$
and a background $SU(2)_R$ gauge field ${G_\m}_A^{\;\; B}$.
We are looking for a solution that satisfies the following reality condition:
\be\label{eq:reality}
(\xi_{\alpha A})^\dagger=\xi^{A \alpha}=\epsilon^{\alpha\beta}\epsilon^{AB}\xi_{\beta B}, \quad
(\chir\xi_{\dot\alpha A})^\dagger=\chir\xi^{A \dot\alpha}=\epsilon^{\dot\alpha\dot\beta}\epsilon^{AB}\chir\xi_{\dot\beta B}.
\ee
If the spinors $\xi'_A,\chir\xi'_A$ on the r.h.s. of (\ref{eq:1}) are orthogonal to $\xi_A,\chir\xi_A$, i.e. if
\be\label{eq:orthogonality}
\xi^{\prime A\alpha}\xi_{A\alpha}=0, \qquad \chir\xi^{\prime A\alpha}\chir\xi_{A\alpha}=0,
\ee
then they can be parametrized as follows
\be
\xi'_A=-i S^{\kappa\lambda}\s_{\kappa\lambda}\xi_A, \qquad
\chir\xi'_A=-i \chir S^{\kappa\lambda}\chir\s_{\kappa\lambda}\chir\xi_A,
\ee
where $S$, $\chir S$ are respectively anti self-dual and self-dual tensors.
If this happens, equation (\ref{eq:1}) can be written as
\be
\begin{aligned}\label{eq:1bis}
D_\m \xi_A &= - T^{\kappa\lambda}\s_{\kappa\lambda}\s_\m \chir{\xi}_A  - \chir S^{\kappa\lambda} \s_\m \chir\s_{\kappa\lambda}\chir\xi_A, \\
D_\m \chir\xi_A &= - \chir T^{\kappa\lambda}\chir\s_{\kappa\lambda}\chir\s_\m \xi_A  - S^{\kappa\lambda} \chir\s_\m \s_{\kappa\lambda}\xi_A.
\end{aligned}
\ee


\subsection{Twisting solutions}

\subsubsection{Witten twisting solutions}

The problem of finding solutions to (\ref{eq:1}) simplifies a lot if we turn on the background $SU(2)_R$
gauge field ${G_\m}_A^{\;\; B}$ in equation (\ref{eq:Dcov}).
Turning on only the $U(1)_R\subset SU(2)_R$ component ${G_\m}_A^{\;\;B}=G_\m\, {\s_3}_A^{\;\;B}$
means twisting the Euclidean rotation group as $SO(4)'\subset SO(4)\times U(1)_R$.
The twisted theory is obtained gauging $SO(4)'$ by the spin connection.
In this way the spinors of the untwisted theory become sections of different bundles.

We now derive the simplest solution of (\ref{eq:1}) performing the following twist:
\be\label{eq:Wittentwist}
{G_\m}_A^{\;\;B}=G_\m\, {\s_3}_A^{\;\;B}, \qquad G_\m=-\frac{1}{2}(\omega_\m+\omega'_\m),
\ee
where $\omega_\m,\omega'_\m$ are the components of the spin connection (see appendix \ref{App:metric}, equation (\ref{Spin-Conn}))
\be\label{eq:spinconn}
\begin{aligned}
&\omega_\m=\omega_\m^{12}=-2i \omega_{\m 1\bar 1}, \\
&\omega'_\m=\omega_\m^{34}=-2i \omega_{\m 2\bar 2}. \\
\end{aligned}
\ee
The right hand side of equation (\ref{eq:1}) becomes
\be\label{eq:twisted1}
\begin{aligned}
D_\m\xi_1 &=(\de_\m-i \omega_\m \Pm -i \omega'_\m \Pp)\xi_1,\\
D_\m\xi_2 &=(\de_\m+i \omega_\m \Pp -i \omega'_\m \Pm)\xi_2, \\
D_\m\chir\xi_1 &=(\de_\m-i \omega_\m \Pm -i \omega'_\m \Pp)\chir\xi_1,\\
D_\m\chir\xi_2 &=(\de_\m+i \omega_\m \Pp -i \omega'_\m \Pm)\chir\xi_2,
\end{aligned}
\ee
where $\Pp$ and $\Pm$ are respectively the projectors in the first and in the second component of the two components Weyl spinor
\be
P_\pm:=\frac{\Id \pm\s_3}{2}.
\ee
It is easy to check that
\be\label{eq:WittenSol}
\xi_A=0, \qquad
\chir\xi_1=\left( \begin{array}{c} a \\ 0 \end{array} \right), \quad
\chir\xi_2=\left( \begin{array}{c} 0 \\ \bar a \end{array} \right), \quad a \in \mathbb{C}
\ee
(where the bar over $a$ means taking the complex conjugate) is a solution to the equations (\ref{eq:1}), (\ref{eq:2}) and (\ref{eq:reality}) with
\be
T_{\m\n}=\chir T_{\m\n}=0, \qquad \xi'_A=\chir\xi'_A=0, \qquad M=0.
\ee

The theory invariant under the supersymmetry generated by the   
solution (\ref{eq:WittenSol}) coincides with Witten's topologically twisted version of Super Yang-Mills \cite{Witten:1988ze}. The corresponding
path integral localizes on the moduli space of anti-instantons on $S^2\times S^2$. The integration over this moduli space is however a difficult task, and can be simplified 
further by exploiting the isometry of the base manifold $S^2\times S^2$ 
by considering a new supersymmetry generator which closes on a $U(1)$ isometry. To this end, one has to find  
another set of solutions where $\xi_A \neq 0$, as we will show in the next subsection.

\subsubsection{Equivariant twisting solutions}
\label{equiv-twisted-sol}

We will follow the procedure described in \cite{Closset:2014pda} to obtain a
more general solution for the twist (\ref{eq:Wittentwist}).
This procedure is actually available more in general for a generic Riemannian four-manifold admitting a $U(1)$ isometry.
We report the general result in appendix \ref{app:GenSol}.

We would like to find a supersymmetry generator which squares on an isometry of the base manifold, in order to
localize the path integral to its fixed points.
To obtain this we have to turn on the left chirality solution $\xi_A$.
Indeed the vector generating the isometry is proportional to
\be\label{eq:KillVec}
\frac{1}{2}V_a=\chir\xi^A\chir\s_a\xi_A,
\ee
where $\chir\s_a=(i\s_1,i\s_2,i\s_3,\Id)$,
or using complex coordinate in the orthonormal frame
\be\label{eq:sigmacpx}
\begin{aligned}
&\chir \s_1=\frac{1}{2}i\s_1+\frac{1}{2i}i\s_2=i\left( \begin{array}{cc} 0 & 0 \\ 1 & 0 \end{array} \right), \qquad
\chir \s_{\bar 1}=\frac{1}{2}i\s_1-\frac{1}{2i}i\s_2=i\left( \begin{array}{cc} 0 & 1 \\ 0 & 0 \end{array} \right), \\
&\chir \s_2=\frac{1}{2}i\s_3+\frac{1}{2i}\Id=i\left( \begin{array}{cc} 0 & 0 \\ 0 & -1 \end{array} \right), \qquad
\chir \s_{\bar 2}=\frac{1}{2}i\s_3-\frac{1}{2i}\Id=i\left( \begin{array}{cc} 1 & 0 \\ 0 & 0 \end{array} \right).
\end{aligned}
\ee
The vector field that we will consider for $S^2\times S^2$ is
\be\label{Kvec}
\frac{1}{2}V=\ve_1(iz\de_z-i\bar z\de_{\bar z})+\ve_2(iw\de_w-i\bar w\de_{\bar w})
\ee
where $z$ and $w$ are complex coordinates on the two $S^2$s and $\ve_1=1/r_1$, $\ve_2=1/r_2$ their radii 
\footnote{The isometry generated by $V$ is actually a diagonal combination of the two
isometries of the maximal torus $U(1)^2\subset SO(4)$.
To obtain  separately the action of the two $U(1)$
one can consider complexified version of (\ref{Kvec}) with complex parameters $\ve_1$, $\ve_2$.
One can obtain solutions generating such an isometry relaxing the condition of reality of the
spinors (\ref{eq:reality}).}.

We want to find a solution $(\xi_A,\bar\xi_A)$ that satisfies (\ref{eq:KillVec}), where $V$ is as given in (\ref{Kvec}).
Expanding (\ref{eq:KillVec}) and denoting the two components of a Weyl spinor $\psi=(\psi^+,\psi^-)$ we obtain the following four equations
\be\label{eq:VecCond1}
\begin{aligned}
&V_1=2i\chir\xi_2^+\xi_1^+ -2 i\chir\xi_1^+ \xi_2^+, \\
&V_{\bar 1}= 2i\chir\xi_1^- \xi_2^- -2 i\chir\xi_2^-\xi_1^-, \\
&V_2=2i\chir\xi_1^+ \xi_2^- -2 i\chir\xi_2^+\xi_1^-, \\
&V_{\bar 2}= 2i\chir\xi_1^- \xi_2^+ -2 i\chir\xi_2^-\xi_1^+. \\
\end{aligned}
\ee
Let us fix $a=1$ in equation (\ref{eq:WittenSol}),
then we turn on the zero components of the \emph{real} spinors $\xi_A$, $\chir\xi_A$ as
\be\label{eq:SolGen1}
\xi_1=\left( \begin{array}{c} b \\ c \end{array} \right), \quad
\xi_2=\left( \begin{array}{c} -\bar c \\ \bar b \end{array} \right), \quad
\chir\xi_1=\left( \begin{array}{c} 1 \\ d \end{array} \right), \quad
\chir\xi_2=\left( \begin{array}{c} -\bar d \\ 1 \end{array} \right).
\ee
The covariant derivatives of $b,c,d$ have the following expressions due to the twist (\ref{eq:Wittentwist})
\be\label{eq:section}
\begin{aligned}
&D_\m b =(\de_\m - i\omega'_\m)b, \\
&D_\m c =(\de_\m - i\omega_\m)c, \\
&D_\m d =(\de_\m - i\omega_\m-i\omega'_\m)d.
\end{aligned}
\ee
Putting (\ref{eq:SolGen1}) in (\ref{eq:VecCond1}) we obtain the system
\be\label{eq:VecCond2}
\begin{aligned}
&V_1          =2(i\bar c - i \bar d b), \\
&V_{\bar 1}=2(-i c + i d \bar b), \\
&V_2          =2(i\bar b + i \bar d c), \\
&V_{\bar 2}= 2(-i b - i d \bar c), \\
\end{aligned}
\ee
where of course $\overline{(V_1)}=V_{\bar 1}$ and $\overline{(V_2)}=V_{\bar 2}$ due to the reality of the spinor,
(which is equivalent to reality of the vector $V$).
The simplest choice for $b,c,d$ is
\be
b=\frac{i}{2}V_{\bar 2}, \qquad  c=\frac{i}{2}V_{\bar 1}, \qquad d=0.
\ee

Now we have to show that this is actually a solution to equation (\ref{eq:1})
for some values of the background fields.
We can rewrite (\ref{eq:section}) using (\ref{eq:spinconn}) in terms of $V_1$ and $V_2$ as
\be
\begin{aligned}
&(\de_\m+i\omega_\m)V_1          =(\de_\m+\omega_{\m 1}^{\phantom{\m 1}1})V_1         =\nabla_\m V_1, \\
&(\de_\m-i\omega_\m)V_{\bar 1}=(\de_\m+\omega_{\m \bar 1}^{\phantom{\m \bar1}\bar1})V_{\bar 1}=\nabla_\m V_{\bar 1}, \\
&(\de_\m+i\omega'_\m)V_1          =(\de_\m+\w_{\m 2}^{\phantom{\m 2}2})V_2         =\nabla_\m V_2, \\
&(\de_\m-i\omega'_\m)V_{\bar 2}=(\de_\m+\w_{\m \bar 2}^{\phantom{\m \bar 2}\bar 2})V_{\bar 2}=\nabla_\m V_{\bar 2}.
\end{aligned}
\ee
Using the properties of Killing vectors and the factorization of the metric
(see equations (\ref{eq:D1form}), (\ref{eq:complexmetric}) in appendix \ref{App:metric}),
it's easy to show that the only non-zero components of $\nabla_\m V_\n$ are
\be
\nabla_z V_{\bar 1}=e_{\bar 1}^{\bar z} \nabla_{z}V_{\bar z}=e_{\bar 1}^{\bar z} \nabla_{[z}V_{\bar z]}
			   =e_{\bar 1}^{\bar z} \de_{[z}V_{\bar z]}
			   =\frac{1}{2}\left(\frac{\de_z V_{\bar z}-\de_{\bar z} V_z}{\sqrt{g_1}}\right)g_1^{1/4}
			   =:\mathcal{H}_1\, g_1^{1/4};
\ee
and similarly\footnote{
$g_1,g_2$ are respectively the determinants of the metric in the first and in the second sphere,
$\sqrt{g_1}:=2g_{z\bar z}$ and $\sqrt{g_2}:=2g_{w\bar w}$.}
\be
\begin{aligned}
&\nabla_{\bar z} V_{1}=-\mathcal{H}_1\, g_{(1)}^{1/4}, \\
&\nabla_{w} V_{\bar 2}=+\mathcal{H}_2\, g_{(2)}^{1/4}, \\
&\nabla_{\bar w} V_{2}=-\mathcal{H}_2\, g_{(2)}^{1/4},
\end{aligned}
\ee
where $\mathcal{H}_1$ and $\mathcal{H}_1$ are proportional to 
the height functions on the first and the second sphere respectively.
Indeed considering the Killing vector
$V=\frac{1}{r_1}(iz\de_z -i\bar z\de_{\bar z})+\frac{1}{r_2}(iw\de_w -i\bar w\de_{\bar w})$
we have
\be
\begin{aligned}
&\mathcal{H}_1:=\frac{\de_z V_{\bar z}-\de_{\bar z} V_z}{2\sqrt{g_{(1)}}}
               =\frac{i}{r_1}\frac{1-|z|^2}{1+|z|^2} = \frac{i}{r_1}\cos\theta_1, \\
&\mathcal{H}_2:=\frac{\de_w V_{\bar w}-\de_{\bar w} V_w}{2\sqrt{g_{(2)}}}
               =\frac{i}{r_2}\frac{1-|w|^2}{1+|w|^2} = \frac{i}{r_2}\cos\theta_2.
\end{aligned}
\ee
Using these facts and recalling the form of the candidate solution
\be\label{eq:SolGen2}
\xi_1=\frac{i}{2}\left( \begin{array}{c}  V_{\bar 2} \\  V_{\bar 1} \end{array} \right), \quad
\xi_2=\frac{i}{2}\left( \begin{array}{c}  V_1 \\ - V_2 \end{array} \right), \quad
\chir\xi_1=\left( \begin{array}{c} 1 \\ 0 \end{array} \right), \quad
\chir\xi_2=\left( \begin{array}{c}  0 \\ 1 \end{array} \right),
\ee
we get the following equations for the  left chirality spinor $\xi_A$:
\be\label{eq:left1}
\begin{array}{ll}
  D_{z}\xi_1=\frac{i}{2}\mathcal{H}_1\,g_{(1)}^{1/4}\left( \begin{array}{c}  0 \\ 1 \end{array} \right), \qquad
&D_{z}\xi_2=0,  \\
  D_{w}\xi_1=\frac{i}{2}\mathcal{H}_2\,g_{(2)}^{1/4}\left( \begin{array}{c}  1 \\ 0 \end{array} \right), \qquad
&D_{w}\xi_2=0,  \\
  D_{\bar z}\xi_1=0, \qquad
&D_{\bar z}\xi_2=-\frac{i}{2}\mathcal{H}_1\,g_{(1)}^{1/4}\left( \begin{array}{c}  1 \\ 0 \end{array} \right), \\
  D_{\bar w}\xi_1=0, \qquad
&D_{\bar w}\xi_2=\frac{i}{2}\mathcal{H}_2\,g_{(2)}^{1/4}\left( \begin{array}{c}  0 \\ 1 \end{array} \right). \\
\end{array}
\ee
This can be rewritten in a clever way as
\be\label{eq:left2}
\begin{aligned}
&D_{z}\xi_A=-\frac{1}{2}\mathcal{H}_1\,\s_{z}\,\chir\xi_A,  \\
&D_{w}\xi_A=-\frac{1}{2}\mathcal{H}_2\,\s_{w}\,\chir\xi_A,   \\
&D_{\bar z}\xi_A=+\frac{1}{2}\mathcal{H}_1\,\s_{\bar z}\,\chir\xi_A,  \\
&D_{\bar w}\xi_A=+\frac{1}{2}\mathcal{H}_2\,\s_{\bar w}\,\chir\xi_A,
\end{aligned}
\ee
where $\s_z=e_z^1\,\s_1=g_{(1)}^{1/4}\,\s_1$ etc., and $\s_1,\s_{\bar 1},\s_2,\s_{\bar 2}$
are defined analogously to (\ref{eq:sigmacpx}).

The last thing to do now is to express the background $T,\chir S$ of (\ref{eq:1bis})
in terms of $\mathcal{H}_1,\mathcal{H}_2$
since we already know
\be
D_\m\chir\xi_A=0 \quad \Rightarrow \quad \chir T,S=0.
\ee
Staring at (\ref{eq:left2}) one can notice that we need to associate $\mathcal{H}_1$ to the coordinates $z,\bar z$ of the first sphere and
$\mathcal{H}_2$ to the coordinates $w,\bar w$ of the second sphere.
To reproduce this in (\ref{eq:1bis}) we need the combinations $T^{\kappa\lambda}\s_{\kappa\lambda}$, $\chir S^{\kappa\lambda}\chir\s_{\kappa\lambda}$
to be proportional to $\s_3$, since this matrix has the property
\be
\{\s_3,\s_{z}\}=\{\s_3,\s_{\bar z}\}=0, \qquad [\s_3,\s_{w}]=[\s_3,\s_{\bar w}]=0.
\ee
Therefore the only possibility is
\be\label{eq:backG1}
T=t\big(\w(1)-\w(2)\big), \qquad \chir S=s\big(\w(1)+\w(2)\big),
\ee
where $t$ and $s$ are two real scalar functions and $\w(1),\w(2)$ are respectively the volume forms on the first
and on the second sphere.
Indeed from (\ref{eq:backG1}) we have
\be
T^{\kappa\lambda}\s_{\kappa\lambda}=4it\s_3, \qquad  \chir S^{\kappa\lambda}\chir\s_{\kappa\lambda}=4is\s_3.
\ee
Inserting these two in equations (\ref{eq:1bis}) and using the further property
\be
\s_{z}\s_3=\s_{z}, \quad \s_{\bar z}\s_3=-\s_{\bar z}, \quad \s_{w}\s_3=\s_{w}, \quad \s_{\bar w}\s_3=-\s_{\bar w},
\ee
we obtain
\be
\begin{aligned}
&D_{z}\xi_A=4i(t-s)\s_{z}\chir\xi_A, \\
&D_{\bar z}\xi_A=4i(-t+s)\s_{\bar z}\chir\xi_A, \\
&D_{w}\xi_A=4i(-t-s)\s_{w}\chir\xi_A, \\
&D_{\bar w}\xi_A=4i(t+s)\s_{\bar w}\chir\xi_A.
\end{aligned}
\ee
Finally comparing with (\ref{eq:left2}) we get
\be
t-s=\frac{i}{8}\mathcal{H}_1, \quad t+s=-\frac{i}{8}\mathcal{H}_2 \quad \Rightarrow \quad
t=\frac{i}{16}(\mathcal{H}_1-\mathcal{H}_2), \quad s=-\frac{i}{16}(\mathcal{H}_1+\mathcal{H}_2).
\ee

It remains to evaluate the background field $M$ in (\ref{eq:2}).
From the second of (\ref{eq:2}) it trivially follows
\be\label{eq:backM}
M=0
\ee
due to $D_\m \chir \xi_A=0$ and $\chir T=0$.
It is of course possible to check this result also using the first of (\ref{eq:2}) and inserting the values of $D_\m \xi_A$ and $T$.

\subsubsection{Summary of the results}
We summarize here the results of the previous subsection
\be\label{eq:ResumeSol}
\xi_1=\frac{i}{2}\left( \begin{array}{c}  V_{\bar 2} \\  V_{\bar 1} \end{array} \right), \quad
\xi_2=\frac{i}{2}\left( \begin{array}{c}  V_1 \\ - V_2 \end{array} \right), \quad
\chir\xi_1=\left( \begin{array}{c} 1 \\ 0 \end{array} \right), \quad
\chir\xi_2=\left( \begin{array}{c}  0 \\ 1 \end{array} \right),
\ee
satisfying
\be
\begin{aligned}\label{eq:1bis-twist}
D_\m \xi_A &= - T^{\kappa\lambda}\s_{\kappa\lambda}\s_\m \chir{\xi}_A  - \chir S^{\kappa\lambda} \s_\m \chir\s_{\kappa\lambda}\chir\xi_A, \\
D_\m \chir\xi_A &=0,
\end{aligned}
\ee
with
\be\label{eq:backG2}
T=t\big(\w(1)-\w(2)\big), \quad \chir S=s\big(\w(1)+\w(2)\big),
\ee
and
\be
t=\frac{i}{16}(\mathcal{H}_1-\mathcal{H}_2), \quad s=-\frac{i}{16}(\mathcal{H}_1+\mathcal{H}_2), \quad
\mathcal{H}_1=\frac{\de_z V_{\bar z}-\de_{\bar z} V_z}{2\sqrt{g_{(1)}}}, \quad
\mathcal{H}_2=\frac{\de_w V_{\bar w}-\de_{\bar w} V_w}{2\sqrt{g_{(2)}}}.
\ee
In polar coordinates of the two spheres these are
\be\label{susy-polar}
\xi_1=-\frac{1}{2}\left( \begin{array}{c} \sin\theta_2 \\ \sin\theta_1 \end{array} \right), \quad
\xi_2=\frac{1}{2}\left( \begin{array}{c} \sin\theta_1 \\ -\sin\theta_2 \end{array} \right), \quad
\chir\xi_1=\left( \begin{array}{c} 1 \\ 0 \end{array} \right), \quad
\chir\xi_2=\left( \begin{array}{c}  0 \\ 1 \end{array} \right),
\ee
and
\be
t=\frac{1}{16}\Big(-\frac{\cos\theta_1}{r_1}+\frac{\cos\theta_2}{r_2}\Big), \qquad
s=\frac{1}{16}\Big(\frac{\cos\theta_1}{r_1}+\frac{\cos\theta_2}{r_2}\Big).
\ee
The square norms of the spinors are
\be\label{sq-spinors}
\xi^2:=\xi^A\xi_A=\frac{1}{2}(\sin^2\theta_1+\sin^2\theta_2),\qquad
\chir\xi^2:=\chir\xi_A\chir\xi^A=2.
\ee
Instead in complex coordinates
\be
\xi_1=-\left( \begin{array}{c} \frac{w}{1+|w|^2} \\ \frac{z}{1+|z|^2} \end{array} \right), \quad
\xi_2=\left( \begin{array}{c} \phantom{-}\frac{\bar z}{1+|z|^2} \\ -\frac{\bar w}{1+|w|^2} \end{array} \right), \quad
\chir\xi_1=\left( \begin{array}{c} 1 \\ 0 \end{array} \right), \quad
\chir\xi_2=\left( \begin{array}{c}  0 \\ 1 \end{array} \right),
\ee
and
\be
t=\frac{1}{16}\left(-\frac{1}{r_1}\frac{1-|z|^2}{1+|z|^2}+\frac{1}{r_2}\frac{1-|w|^2}{1+|w|^2}\right), \qquad
s=\frac{1}{16}\left(\frac{1}{r_1}\frac{1-|z|^2}{1+|z|^2}+\frac{1}{r_2}\frac{1-|w|^2}{1+|w|^2}\right).
\ee




\section{Partition function on $S^2\times S^2$ }
\label{twist}

In this section we proceed to the computation of the partition function of $\N=2$ SYM theory on  $S^2\times S^2$.
At the end of the section we will present the extension of this result in presence of matter fields 
in the (anti)fundamental representation.

The strategy we follow consists of performing a change of variables in the path integral to an equivariant
extension of the Witten's topologically twisted theory \cite{Witten:1988ze}.
As we will see this maps the supersymmetry algebra to an equivariant BRST algebra which is 
the natural generalization of the supersymmetry algebra 
of the Nekrasov $\Omega$-background \cite{Nekrasov:2002qd,Baulieu:2005bs}
on $S^2\times S^2$.
Since this is a toric manifold, the partition function reduces to copies of the Nekrasov partition functions, glued together in a way that will be explained below. The result we obtain is indeed in agreement with the one conjectured by Nekrasov in \cite{Nekrasov:2003vi}
for toric compact manifolds.

\subsection{Change to twisted variables}

The starting supersymmetry algebra for the vector multiplet is (\ref{Q-alg})
and the square of the supersymmetry action is (\ref{Qsq-alg}),
where the parameters $w=0$ and $\Theta=0$ due to the orthogonality of our solution (\ref{eq:orthogonality}).

Now we are going to make a change of variables in the supersymmetry algebra
that will simplify the localization procedure in the path integral.
We re-organize the eight components of the fermions $\lambda_{A\alpha}, \chir\lambda_{A}^{\dot\alpha}$
as a fermionic scalar $\eta$, a vector $\Psi^\m$ and a self dual tensor $\chi^{(+)\m\n}$:
\be\label{eq:changevar}
\begin{aligned}
\eta&:=-i(\xi^A\lambda_A+\chir\xi^A\chir\lambda_A), \\
\Psi_\m&:=i(\xi^A\s_\m\chir\lambda_A - \chir\xi^A\chir\s_\m\lambda_A), \\
\chi^{+}_{\m\n}&:=2\chir\xi^{A}\chir\sigma_{\m\n}\chir\xi^{B}(\chir\xi_{A}\chir\lambda_{B}-\xi_{A}\lambda_{B}).
\end{aligned}
\ee

We also redefine the scalars in a suitable way to simplify the supersymmetry algebra:
\be
\begin{aligned}
\chir\Phi:=&\phi-\chir\phi\\
\Phi:=&2i\chir\xi^2\phi+2i\xi^2\chir\phi\\
B^+_{\m\n}:=&\,2(\chir\xi^2)^2(F_{\m\n}^+ +8\phi \chir T_{\m\n} - 8\chir\phi \chir S_{\m\n}) \\
           &-(\chir\xi^A\chir \sigma_{\m\n}\chir\xi^B)   
           (\xi_A\s^{\kappa\lambda}\xi_B)(F_{\kappa\lambda}+8\chir\phi T_{\kappa\lambda} - 8\phi S_{\kappa\lambda}) \\
           &-4\chir\xi^2V_{[\m} D_{\n]}^+\chir\Phi
           +\frac{1}{2}(\xi^2+\chir\xi^2) (\chir\xi^A\chir\sigma_{\m\n}\chir\xi^B) D_{AB}.
\end{aligned}
\ee
where $\xi^2$ and $\chir\xi^2$ are the square norms of the spinors (\ref{sq-spinors}). 

The inverse of the relation (\ref{eq:changevar}) is given by
\be\label{eq:Inverse-changevar}
\begin{aligned}
&\lambda_{A}=\frac{1}{\xi^2+\chir\xi^2}
                   (i\xi_A\eta-i\s^\mu\chir\xi_A\Psi_\mu+\xi^B\Xi_{BA}) \\
&\chir\lambda_{A}=\frac{1}{\xi^2+\chir\xi^2}
                        (-i\chir\xi_A\eta-i\chir\s^\mu\xi_A\Psi_\mu+\chir\xi^B\Xi_{BA})
\end{aligned}
\ee
where
\be
\Xi_{AB}=\frac{1}{2(\chir\xi^2)^2}\chir\xi_A\chir\s^{\m\n}\chir\xi_B\chi^+_{\m\n}.
\ee
It is immediate to verify the relation (\ref{eq:Inverse-changevar}) by inserting it back in (\ref{eq:changevar});
or conversely by inserting (\ref{eq:changevar}) in (\ref{eq:Inverse-changevar}) and using the following non-trivial spinor identity
\be
(\psi^B\cdot\lambda_B)\xi_A-(\psi_A\cdot\lambda_B)\xi^B-(\psi_B\cdot\lambda_A)\xi^B=0,
\ee
for two-components spinors $\psi,\lambda$,$\xi$.

A comment is now important: this change of variables is everywhere invertible.
Indeed, its Jacobian is given by
\be
\begin{aligned}
{\cal J}&=\frac{{\cal J}_{bos}}{{\cal J}_{ferm}}=1
\\
{\cal J}_{bos}&={\cal J}_{ferm}\sim \left(\chir\xi^2+\xi^2\right)^4(\chir\xi^2)^3.
\end{aligned}
\ee

Notice that this
change of variables is everywhere well defined due to the nature of the
solution derived in section \ref{equiv-twisted-sol}.
Indeed, neither of the factors $\chir\xi^2$ and 
$\xi^2+\chir\xi^2$ in the Jacobian ever vanish.

The supersymmetry algebra in terms of the new variables, computed from (\ref{Q-alg}) and (\ref{Qsq-alg}), is
\be\label{eq:newSUSY1}
\begin{aligned}
&\Q A=\Psi, &\quad &\Q\Psi=i\iota_V F + D\Phi, &\quad& \Q \Phi= i \iota_V\Psi,\\
&\Q\chir\Phi=\eta, &\quad &\Q\eta=i\,\iota_VD\chir\Phi+i[\Phi,\chir\Phi], \\
&\Q\chi^+=B^+, &\quad &\Q B^+=i\mathcal{L}_V\chi^++i[\Phi,\chi^+].
\end{aligned}
\ee
These are the equivariant extension of the twisted supersymmetry considered in \cite{Witten:1988ze}
and we finally got rid of all the indices in our formulas by passing to the differential form notation.

In \eqref{eq:newSUSY1} $\iota_V$ is the contraction with the vector $V$ and $\mathcal{L}_V=D\iota_V+\iota_V D$ is the covariant Lie derivative.

Let us notice that the supercharge \eqref{eq:newSUSY1} manifestly satisfies $\Q^2=i \mathcal{L}_V +\delta^{gauge}_\Phi$.
There is still a consistency condition on the last line, that is the action has to preserve the self-duality 
of $B^+$ and $\chi^+$. This is satisfied iff $L_V \star = \star L_V$, where $\star$ is the Hodge-$\star$ and $L_V=d\iota_V+\iota_V d$
is the Lie derivative. This condition coincides with the requirement that $V$ is an isometry of the four manifold.
Therefore, we have proved that for any four-manifold with a $U(1)$ isometry, once the $R$-symmetry bundle is chosen to fit the equivariant twist, 
there is a consistent realization of the corresponding $\N=2$ supersymmetry algebra\footnote{In terms of the vector field, the Jacobian factors
above read ${\cal J}_{bos}={\cal J}_{ferm}\sim (2+\frac{1}{8}V^2)$ which is positive.},
explicit formulae for the generators of supersymmetry and background fields in this general case are reported in appendix \ref{app:GenSol}.

\subsection{Localizing action and fixed points}

In terms of the new variables (\ref{eq:changevar}), we consider the following supersymmetric Lagrangian
\be\label{nonchanome}
L=\frac{i\tau}{4\pi}\Tr F\wedge F + \omega\wedge\Tr F +\Q \mathcal{V}
\ee
where $\tau$ is the complexified coupling constant, $\omega\in H^2\left(S^2\times S^2,{\mathbb R}\right)$ and
\be\label{GFFform}
\begin{aligned}
\mathcal{V}=&-\text{Tr}\big[\chi^+\wedge\star F
          + \star i \chir\Phi(-\star D\star \Psi+ \mathcal{L}_V\eta)
          + \star\eta (i\,\mathcal{L}_V\chir\Phi+i[\Phi,\chir\Phi])^\dagger\big] \\
	  &-\text{Tr}\big[\chi^+\big]\wedge\star\text{Tr}\big[B^+\big].
\end{aligned}
\ee

Proceeding to discuss the localization of the gauge field,
we will split the calculation between the $u(1)$ and the $su(N)$ sector
which must be differently treated.
This is due to the fact that we want to allow gauge vector bundles with non trivial and
unrestricted first Chern class $c_1=\frac{1}{2\pi}Tr F$. 
The usual $\delta$-gauge fixing $F^+=0$ in the whole $u(N)$
Lie algebra would be then incompatible with the previous request.
Therefore we split the gauge fixing of the two sectors with the additional term in the last line of \eqref{GFFform}, keeping 
a Gaussian gauge fixing in the $u(1)$ sector and a $\delta$-gauge fixing in the $su(N)$ sector.
If the manifold is K\"ahler, then an equivalent procedure would be to localize on Hermitian-Yang-Mills connections, namely those
satisfying the equation $\omega\wedge F= c\,\, \omega\wedge\omega \Id$ and $F^{(2,0)}=0$, where $\omega$ is the K\"ahler form and $c$ is a constant.

We look at the fixed points of the supersymmetry (\ref{eq:newSUSY1}).
On setting the fermions to zero, the fixed points of the supercharge read
\be\label{BPScond1}
\begin{aligned}
&\iota_VD\chir\Phi+[\Phi,\chir\Phi]=0, \\
&i\iota_V F + D\Phi=0.
\end{aligned}
\ee
The integrability conditions of the second equation are
\be\label{BPS-int}
\begin{aligned}
&\iota_V D\Phi=0, \\
&\mathcal{L}_V F=[F,\Phi].
\end{aligned}
\ee

We choose the following reality condition for the scalars fields
$\chir\Phi=-\Phi^\dagger$, then the first of \eqref{BPScond1} splits as 
\be
\iota_VD\chir\Phi=0 \quad {\rm and} \quad [\Phi,\chir\Phi]=0.
\ee
which imply that $\Phi$ and $\chir\Phi$ lie in the same Cartan subalgebra.
Moreover, since we consider $F^\dagger=F$, we can split
similarly the second of (\ref{BPS-int}),
obtaining that also the curvature lies along the Cartan subalgebra
\be\label{F-Cartan}
[F,\Phi]=[F,\bar\Phi]=0.
\ee
Therefore the second of (\ref{BPScond1}) can be rewritten as
\be\label{moment-map}
\iota_V F=i\,d\Phi
\ee
since the extra term $[A,\Phi]$ is different from zero only outside
the Cartan subalgebra.
(\ref{moment-map}) means that $\Phi$ is the moment map
for the action of $V$ on $F$.

%

The gauge fixing condition comes by
integrating out the auxiliary field $B^+$
from (\ref{GFFform}).
As anticipate we obtain different gauge conditions
for the $u(1)$ and $su(N)$ sector\footnote{
%
%
For the $u(1)$ sector,
define $f:=F_{u(1)}$ and $b^+:=B^+_{u(1)}.$
From (\ref{GFFform}) we have
$\mathcal{QV}_{u(1)}=- b^+\wedge\star f^{+} - b^+\wedge \star b^{+}.$
Integrating out $b^+$ we get the condition $b^+=-f^+/2$ and inserting this back, we obtain
$\mathcal{QV}_{u(1)}= \frac{1}{4}f^+\wedge\star f^{+}=\frac{1}{8}(f\wedge f+f\wedge\star f)$
that give the equation of motion $d\star f=0$.
}
%
%
\be
d\star(F_{u(1)})=0, \qquad (F_{su(N)})^+=0.
\ee
In particular $d\star F=0$ in the whole $u(N)$.
This, together with the Bianchi identity $d F=0$
and the fact that $F$ lies in the Cartan subalgebra of $u(N)$
(\ref{F-Cartan}),
implicates that the curvature must be a harmonic 2-form
with values on the Cartan subalgebra and integer periods.
Namely for each elements in the Cartan subalgebra,
labeled by $\alpha=1,\dots,N$
\be
(c_1)_\alpha\equiv\frac{i F_\alpha}{2\pi}\in H^2(S^2\times S^2,\mathbb{Z}).
\ee
So
\be\label{F-u(N)}
\frac{i F_\alpha}{2\pi}=m_\alpha\,\omega(1)+n_\alpha\, \omega(2),
                         \qquad m_\alpha,n_\alpha \in \mathbb{Z}.
\ee
where a basis of normalized harmonic 2-forms for $S^2\times S^2$
is given by
\be
\omega(i)=\frac{1}{4\pi}\sin\theta_i d\theta_i\wedge d\varphi_i
\quad  i=1,2.
\ee

Replacing this expression for $F$ in (\ref{moment-map})
we get
\be
\iota_V (m_\alpha\,\omega(1)+n_\alpha\, \omega(2))=-2\,d\Phi_\alpha
\ee
Since $S^2\times S^2$ is simply connected
the closed 2-forms $\iota_V\omega(1)$, $\iota_V\omega(2)$
are also exact and the equation (\ref{F-u(N)}) can be integrated
\be
\iota_V\omega(1)=2\,dh_1, \qquad \iota_V\omega(2)=2\,dh_2,
\ee
the solution being given respectively
by the height functions on the two spheres\footnote{
The vector generated by the supersymmetry generators (\ref{susy-polar})
is $V=2\ve_1\de_{\varphi_1}+2\ve_2\de_{\varphi_2}$
}
\be
h_1=\ve_1\cos\theta_1, \qquad h_2=\ve_2\cos\theta_2.
\ee
Finally, integrating equation (\ref{F-u(N)}) we obtain
\be\label{Phi-shift}
\Phi_\alpha=-m_\alpha\, h_1-n_\alpha \, h_2+a_\alpha,
\qquad m_\alpha, n_\alpha \in\mathbb{Z}.
\ee
where $a_\alpha$ are integration constants.
On top of fluxes, the complete solution contains also point-like instantons located at the 
zeroes of the vector field $V$. These do not contribute to the equations above since $i_VF^{{\rm point}}=0$.

We are then reduced to a sum over point-like instantons and an integration over the constant 
Cartan valued variable $\Phi$.
Let us notice that the above arguments are quite general and apply to more general 
four-manifolds than $S^2\times S^2$.

Before proceeding to the computation, let us notice that on compact manifolds one has to take care
of normalizable fermionic zero modes of the Laplacian,
counted by the Betti numbers. If the manifold is simply connected, as we assume, then
the field $\Psi$ doesn't display such zero modes, while $\eta$ will display one zero mode and $\chi^+$
will display $b_2^+$ of them (for each element in the $su(N)$ Cartan\footnote{The ones in the $U(1)$ are 
gauge fixed as a BRST quartet.}
labeled by $\rho=1,\dots,N-1$).
On $S^2\times S^2$, and in general on any toric manifold, $b_2^+=1$ and therefore the $(\eta,\chi^+)$
zero modes come in pairs, one pair for each element in the $su(N)$ Cartan. 
One can soak-up those zero modes by adding the exact term
$S_\text{zm}= s \Q \int\sum_\rho \bar\Phi_\rho \chi^+_\rho B_\rho$
to the action whose effect, after the integration over the  $(\eta,\chi^+)$
zero mode pairs and the $B$-zero modes, is to insert a derivative with respect to $\bar a$
for each $su(N)$ Cartan element. This reduces the integration over the $a$-plane to a contour integral around
the diagonals where $a_{\alpha}-a_{\beta}=0$.


\subsection{Computation of the partition function}

Due to the results of the previous section, the integration over the instanton moduli space is reduced to
instanton counting. In particular due to supersymmetry, the instanton configurations have to be equivariant under the action of
$U(1)^{2}\times U(1)^N$ which is the maximal torus of the isometry group of $S^2\times S^2$ times the constant gauge transformations.
It is well known that the fixed points are classified by Young diagrams \cite{Nekrasov:2002qd}, so that
for each fixed point we
have to consider a contribution given by the Nekrasov instanton partition function with the proper torus weights.

Let us underline the important difference between the compact and non-compact case, namely that in the former the gluing of Nekrasov partition functions
also involves the integration over the Cartan subalgebra of the gauge group.
This appears as a contour integral as explained at the end of the previous subsection.

We regard the manifold $S^2\times S^2\cong \mathbb{P}^1\times\mathbb{P}^1$ as a complex toric manifold described in terms of four patches.
The  weights $\big(\ve_1^\li,\ve_2^\li\big)$ of the $(\mathbb{C}^*)^2$ torus action in each patch are
\be\label{weights}
\begin{array}{c|rrrr}
\ell      & 1\;       &  2\;       &   3 \;     &   4\;    \\ \hline \vspace{-12pt} \\
\ve_1^\li & \:\ve_1\: & \:-\ve_2\: & \:-\ve_1\:  & \:\ve_2\: \\
\ve_2^\li & \:\ve_2\: & \: \ve_1\: & \:-\ve_2\: & \:-\ve_1\: 
\end{array}
\ee
where in our case $\ve_1=\frac{1}{r_1}>0$ and $\ve_2=\frac{1}{r_2}>0$ are the inverse radii of the two spheres.

The fixed point data on $S^2\times S^2$ will be described in terms of a collection of Young diagrams $\{\vec Y_\ell \}$,
and of integers numbers $\vec m,\vec n$ describing respectively
the $(\mathbb{C}^*)^{N+2}$-invariant point-like instantons in each patch
(localized at the fixed points $p_\ell$)
and the magnetic fluxes of the gauge field on the spheres
which correspond to the first Chern class $c_1(E)$ of the gauge bundle $E$. 
More explicitly, for a gauge bundle with $c_1=n\omega_1+m\omega_2$ and $ch_2=K$, the fixed point data satisfy
\be
n=\sum_{\alpha=1}^N n_\alpha, \qquad
m=\sum_{\alpha=1}^N m_\alpha, \qquad
K= \sum_{\alpha,\ell}\vert Y_\alpha^{(\ell)} \vert.
\ee
The full partition function on $S^2\times S^2$ is given by 
\be\label{Zfull}
Z^{S^2\times S^2}_{\text{full}}(q,z_1,z_2,\ve_1,\ve_2)=
    \sum_{\{\vec m,\vec n\}}\oint_{\tt t} d\vec a\,
    t_1^{m} t_2^{n}\,
    \prod_{\ell=1}^{4} 
    Z^{\mathbb{C}^2}_{\text{full}}(q,\ve_1^\li,\ve_2^\li, \vec{a}^\li)
\ee
where $q=\exp(2\pi i \tau)$ is the gauge coupling, $t_1=z_1^\frac{1}{2}$ and $t_2=z_2^\frac{1}{2}$ are the 
source terms corresponding to $\omega=v_2\omega_1+v_1\omega_2$ in 
\eqref{nonchanome} so that $z_i=e^{2\pi v_i}$ for $i=1,2$.

Moreover,
$\vec a^\li = \{ a^\li_\alpha \}$, $\alpha=1,\ldots,N$
are the v.e.v.'s of the scalar field $\Phi$
calculated at the fixed points $p_\ell$
\be\label{ashift}
 a^\li_{\alpha} = \langle \Phi(p_\ell) \rangle.
\ee  
Explicitly, by (\ref{Phi-shift})
\be\label{a-patches}
\begin{array}{c|l}
\ell      \,&\, \vec a^\li  \\ \hline \vspace{-8pt} \\
1         \,&\, \vec a+\vec m\ve_1+\vec n\ve_2 \\
2         \,&\, \vec a+\vec m\ve_1-\vec n\ve_2 \\
3         \,&\, \vec a-\vec m\ve_1-\vec n\ve_2 \\
4         \,&\, \vec a-\vec m\ve_1+\vec n\ve_2
\end{array}
\ee

The factors appearing in \eqref{Zfull} are the Nekrasov partition functions  
\be\label{ZfullFact}
	Z^{\mathbb{C}^2}_{\text{full}}(q,\ve_1,\ve_2, \vec{a})=
        Z^{\mathbb{C}^2}_{\text{classical}}(q,\ve_1,\ve_2, \vec{a} )
        Z^{\mathbb{C}^2}_{\text{1-loop}}(\ve_1,\ve_2, \vec{a} )
        Z^{\mathbb{C}^2}_{\text{instanton}}(q,\ve_1,\ve_2, \vec{a}) .
\ee
whose explicit expressions we report below.

\subsubsection{Classical action}
Let us first of all consider the contribution to (\ref{Zfull}) of the classical partition function
\be
\prod_{\ell=1}^{4} Z^{\mathbb{C}^2}_{\text{classical}}(\epsilon_1^\li,\epsilon_2^\li, \vec{a}^\li ).
\ee
For each patch this is given by
\be\label{Zcl}
\begin{aligned}
&Z^{\mathbb{C}^2}_{\text{classical}}(\epsilon_1^\li,\epsilon_2^\li, \vec{a}^\li )
=\exp\left[-\pi i \tau \sum_{\alpha=1}^N
\frac{\big(a_\alpha^\li\big)^2}{(\ve_1^\li \ve_2^\li)}\right]
\end{aligned}
\ee
Inserting the values of the equivariant weights (\ref{weights}) and (\ref{a-patches}) we obtain
\be\label{ZclRES}
\prod_{\ell=1}^{4} Z^{\mathbb{C}^2}_{\text{classical}}(\epsilon_1^\li,\epsilon_2^\li, \vec{a}^\li )=
\exp\left[-\pi i\tau\sum_{\alpha=1}^N 8 m_\alpha n_\alpha\right]
=q^{-4\sum_{\alpha=1}^N m_\alpha n_\alpha}           
\ee
with $q=\exp(2i\pi\tau)$.

\subsubsection{One-loop}
The one-loop contribution in (\ref{Zfull}) is given by
\be\label{Z1loop}
Z^{S^2\times S^2}_{\text{1-loop}}(\vec a,\ve_1,\ve_2) 
=
\prod_{\ell=1}^4 Z^{\mathbb{C}^2}_{\text{1-loop}}(\vec a^\li,\ve_1^\li,\ve_2^\li)
=
\prod_{\ell=1}^4
  \exp\bigg[ -\sum_{\alpha\neq\beta} \gamma_{\ve_1^\li,\ve_2^\li}(a^\li_{\alpha\beta})\bigg]
\ee
where $a_{\alpha\beta}^\li:=a^\li_\alpha-a^\li_\beta$.

Inserting the values of the equivariant weights (\ref{weights}) and (\ref{a-patches})
and using the definition of $\gamma_{\ve_1,\ve_2}$ (appendix \ref{App:special-fun} equation (\ref{gamma2}))
we can rewrite the exponent in (\ref{Z1loop}) as
\be\label{Int1}
-\frac{d}{d s}\Big|_{s=0}\frac{1}{\Gamma(s)}
              \int_0^\infty dt\,t^{s-1}\frac{e^{-ta_{\alpha\beta}}}{(1-x)(1-y)}p(x,y),
\ee
where we have defined $x:=e^{-\ve_1 t}$ and $y:=e^{-\ve_2 t}$ and $p(x,y)$ is a polynomial in $x$ and $y$ given by
\be
p(x,y)=x^{-m}y^{-n}-x^{-m}y^{n+1}-x^{m+1}y^{-n}+x^{m+1}y^{n+1}
\ee
where it is understood that $m\equiv m_{\alpha\beta}$ and $n\equiv n_{\alpha\beta}$.
The residues of this polynomial at $x=1$ and $y=1$ are zero,
this means that in those points $p(x,y)$ has zeros which cancel the poles $(1-x)^{-1},(1-y)^{-1}$
in (\ref{Int1}).

If $m=n=0$
\be
p(x,y)=1-y-x+xy=(1-x)(1-y),
\ee
and integrating (\ref{Int1}) we obtain
\be\label{Z1loop-flux0}
Z^{S^2\times S^2}_{\text{1-loop}}(\vec a,\ve_1,\ve_2)\Big|_{m=n=0}=\prod_{\alpha\neq\beta}a_{\alpha\beta}
=\prod_{\alpha>\beta}(-a_{\alpha\beta}^2).
\ee
In general, for every choice of $\{m,n\}$, one can factorize $(1-x)(1-y)$ out of the polynomial
using the expansion $1-x^N=(1-x)\sum_{j=0}^{N-1}x^j$.
Then $p(x,y)$ can be written as follows
\be
p(x,y)=\left\{
\begin{aligned}
 &(1-x)(1-y)\sum_{j=-m}^m\sum_{k=-n}^n x^j y^k                      &\qquad& \text{if $m\ge 0$, $n\ge 0$} \\[0.01cm]
 &(1-x)(1-y)\sum_{j=m+1}^{-m-1}\sum_{k=n+1}^{-n-1} x^j y^k  &\qquad& \text{if $m< 0$, $n< 0$}    \\[0.01cm]
-&(1-x)(1-y)\sum_{j=-m}^m\sum_{k=n+1}^{-n-1} x^j y^k            &\qquad& \text{if $m\ge 0$, $n< 0$} \\[0.01cm]
-&(1-x)(1-y)\sum_{j=m+1}^{-m-1}\sum_{k=-n}^n x^j y^k            &\qquad& \text{if $m< 0$, $n\ge 0$}
\end{aligned}
\right.
\ee
Using this result in (\ref{Int1}) we obtain for fixed $m$ and $n$
the following result:
\be\label{resVec}
\begin{aligned}
&Z^{S^2\times S^2}_{\text{1-loop}}(a,\ve_1,\ve_2,m,n)\Big|_\text{$\alpha\beta$}= \\
&=\left\{
\begin{aligned}
&\prod_{k=-m}^{m}\prod_{j=-n}^{n}
  (a+k\ve_1+j\ve_2)
	&\qquad& \text{if $m\ge0,n\ge 0$} \\
&\prod_{k=m+1}^{-m-1}\prod_{j=n+1}^{-n-1}
  (a+k\ve_1+j\ve_2)
	&\qquad& \text{if $m<0,n<0$} \\
&\prod_{k=-m}^{m}\prod_{j=n+1}^{-n-1}
  (a+k\ve_1+j\ve_2)^{-1}
	&\qquad& \text{if $m\ge 0,n<0$} \\
&\prod_{k=m+1}^{-m-1}\prod_{j=-n}^{n}
  (a+k\ve_1+j\ve_2)^{-1}
	&\qquad& \text{if $m<0,n\ge 0$}
\end{aligned}\right.
\end{aligned}
\ee
where $a\equiv a_{\alpha\beta}$, $m\equiv m_{\alpha\beta}$ and
$n\equiv n_{\alpha\beta}$.

\subsubsection{Instantons}
The instanton contribution in (\ref{Zfull}) is given by
\be
\prod_{\ell=1}^{4} Z^{\mathbb{C}^2}_{\text{instanton}}(\epsilon_1^\li,\epsilon_2^\li, \vec{a}^\li ).
\ee
where $Z^{\mathbb{C}^2}_{\text{instanton}}$ is the Nekrasov partition function defined as follows.

Let $Y=\{\lambda_1\ge\lambda_2\ge\dots\}$ be a Young diagram,
and $Y'=\{\lambda'_1\ge\lambda'_2\ge\dots\}$ its transposed.
$\lambda_i$ is the length of the i-column and $\lambda'_j$ the length of the j-row of $Y$.
For a given box $s=\{i,j\}$ of the diagram we define
respectively the arm and leg length functions
\be\label{leg-arm}
A_Y(s)=\lambda_i-j, \qquad L_Y(s)=\lambda'_j-i.
\ee
and the arm and leg co-length functions
\be\label{colength}
A'_Y(s)=j-1, \qquad L'_Y(s)=i-1.
\ee
The fixed points data for each patch are given by a collection of Young diagrams $\vec Y^\li=\{Y^\li_\alpha\}$.
and the instanton contribution is
\be\label{ZinstC2}
Z^{\mathbb{C}^2}_{\text{instanton}}(\epsilon_1,\epsilon_2, \vec{a} )=\sum_{\{Y_\alpha\}}q^{|\vec Y|}
                                    z_{\text{vec}}(\epsilon_1,\epsilon_2,\vec a,\vec Y)
\ee
where $q=\exp(2i\pi\tau)$ and
\be\label{Zinstvec}
\begin{aligned}
z_{\text{vec}}(\epsilon_1,\epsilon_2,\vec a,\vec Y)=\prod_{\alpha,\beta=1}^{N}\Bigg[
                             &\prod_{s\in Y_\alpha}\left(a_{\alpha\beta}-L_{Y_\beta}(s)\ve_1+(A_{Y_\alpha}(s)+1)\ve_2\right)\\
                             &\prod_{t\in Y_\beta}\left(a_{\alpha\beta}+(L_{Y_\beta}(t)+1)\ve_1-A_{Y_\alpha}(t)\ve_2\right)
\Bigg]^{-1}.
\end{aligned}
\ee

\subsection{Adding matter fields}

The above formulae are easily modified in presence of matter fields. In the following we discuss the contribution of matter
in the (anti)fundamental representation, which will be used in the last Section when comparing with Liouville gravity.
The contribution to the classical action is vanishing, so we concentrate on one-loop and instanton terms.
 
\subsubsection{One-loop}

When considering matter one has to modify the formula for the one-loop partition function (\ref{Z1loop})
as
\be\label{Z1loop-mat}
Z^{\mathbb{C}^2}_{\text{1-loop}}(\vec a^\li,\ve_1^\li,\ve_2^\li)=
\exp\left[ -\sum_{\alpha\neq\beta} \gamma_{\ve_1^\li,\ve_2^\li}\big(a^\li_{\alpha\beta}\big)
           +\sum_{f=1}^{N_f} \sum_{\rho\in R_f}
            \gamma_{\ve_1^\li,\ve_2^\li}\Big(a^\li_\rho+\mu_f-\frac{Q^\li}{2}\Big)\right]
\ee
where 
$\rho$ are the weights of the representation $R_f$ of the hypermultiplet with mass $\mu_f$ and $R$-charge one,
while $Q^\li:=\ve_1^\li+\ve_2^\li$.

The computation goes as in the previous section,
the additional contribution for each hypermultiplet in the exponential being
\be\label{Int1-mat}
\frac{d}{d s}\Big|_{s=0}\frac{1}{\Gamma(s)}
              \int_0^\infty dt\,t^{s-1}\frac{e^{-t(a_{\rho}+\mu_f+\frac{Q}{2})}}{(1-x)(1-y)}p(x,y).
\ee
where again $x:=e^{-\ve_1 t}$ and $y:=e^{-\ve_2 t}$, with
\be
p(x,y)=x^{-m}y^{-n}-x^{-m}y^{n}-x^{m}y^{-n}+x^{m}y^{n}.
\ee
Eventually, each hypermultiplet contributes as
\be\label{resHyp}
\begin{aligned}
&Z^{S^2\times S^2}_{\text{1-loop}}(a,\ve_1,\ve_2,m,n,\mu_f)\Big|_{\text{hyp},f,\rho}= \\
&=\left\{
\begin{aligned}
&1 & \qquad& \text{if $m\cdot n=0$} \\
&\prod_{k=-m}^{m-1}\prod_{j=-n}^{n-1}
  (a+\mu_f+\frac{Q}{2}+k\ve_1+j\ve_2)^{-1}
	&\qquad& \text{if $m>0,n>0$} \\
&\prod_{k=m}^{-m-1}\prod_{j=n}^{-n-1}
  (a+\mu_f+\frac{Q}{2}+k\ve_1+j\ve_2)^{-1}
	&\qquad& \text{if $m<0,n<0$} \\
&\prod_{k=-m}^{m-1}\prod_{j=n}^{-n-1}
  (a+\mu_f+\frac{Q}{2}+k\ve_1+j\ve_2)
	&\qquad& \text{if $m>0,n<0$} \\
&\prod_{k=m}^{-m-1}\prod_{j=-n}^{n-1}
  (a+\mu_f+\frac{Q}{2}+k\ve_1+j\ve_2)
	&\qquad& \text{if $m<0,n>0$}
\end{aligned}\right.
\end{aligned}
\ee
where $a\equiv a_\rho$, $m\equiv m_\rho$ and $n\equiv n_\rho$. 

\subsubsection{Instantons}

The modification of the instanton partition function due to the presence of matter in the (anti)fundamental
representation is
\be\label{ZinstC2matter}
Z^{\mathbb{C}^2}_{\text{instanton}}(\epsilon_1,\epsilon_2, \vec{a} )=\sum_{\{Y_\alpha\}}q^{|\vec Y|}
                                    z_{\text{vec}}(\epsilon_1,\epsilon_2,\vec a,\vec Y)
                   \prod_{f=1}^{N_F}z_{\text{(anti)fund}}(\epsilon_1,\epsilon_2,\vec a,\vec Y,\mu_f)
\ee
where
\be
\begin{aligned}
&z_{\text{fund}}(\epsilon_1,\epsilon_2,\vec a,\vec Y,\mu_f)=
        \prod_{\alpha}\prod_{s\in Y_\alpha}\left(a_\alpha+L'(s)\ve_1+A'(s)\ve_2+Q-\mu_f\right) \\
&z_{\text{antifund}}(\epsilon_1,\epsilon_2,\vec a,\vec Y,\mu_f)=
        \prod_{\alpha}\prod_{s\in Y_\alpha}\left(a_\alpha+L'(s)\ve_1+A'(s)\ve_2+\mu_f\right)
\end{aligned}
\ee
where $L'(s)$ and $A'(s)$ are the co-length functions defined in (\ref{colength}).

\section{Liouville Gravity}
\label{MLG}

We now proceed to the discussion of a possible two-dimensional Conformal Field Theory (CFT) interpretation of our results,
prompted by AGT correspondence. We focus on the $N=2$ case.
A natural viewpoint to start with is the calculation of the expected central charge via reduction of the 
anomaly polynomial of two M5-branes theory \cite{Bonelli:2009zp,Alday:2009qq}.
Upon compactification on the manifold $S^2\times S^2\times \Sigma$, the central charge of the resulting two-dimensional 
CFT on $\Sigma$ is easily computed via localization formulae from the weights of the $U(1)^2$ torus action, see Table \eqref{weights},
to be
\be
\sum_{\ell=1}^4 \left(1+6\left( b^{(\ell)} + \frac{1}{b^{(\ell)}} \right)^2\right) = 52 = 26 + 26
\label{26}
\ee
were $ b^{(\ell)}=\sqrt{\frac{\ve_1^{(\ell)}}{\ve_2^{(\ell)}}}$. Notice that in passing from one patch to the other only one of
the epsilons change sign so that from real $b$ one passes to imaginary one and viceversa. This will play a relevant role
in the subsequent discussion.
It was observed in \cite{Bershtein:2013oka} that \eqref{26} suggests a link to Liouville gravity.
In the following we will show that indeed three-point number and conformal blocks of this CFT arise as building blocks
of the supersymmetric partition function of $\N=2$ $U(2)$ gauge theory on $S^2\times S^2$.

Liouville Gravity (LG) \cite{Zamolodchikov:2005fy,0510214} 
is a well-known two-dimensional theory of quantum gravity
composed of three CFT sectors
\begin{enumerate}
\item \textbf{Liouville theory sector},
which has a central charge
\be
c_L=1+6Q^2, \qquad Q=b+b^{-1},
\ee
and a continuous family of primary fields parametrized by a complex parameter $\alpha$ as
\be\label{eq:exp-op}
V_\alpha=e^{2\alpha\varphi(x)}
\ee
with conformal dimension
\be
\Delta^L_\alpha=\alpha(Q-\alpha).
\ee

\item \textbf{Matter sector}, a generalized CFT
with central charge
\be
c_M=1-6q^2, \quad q=b^{-1}-b.
\ee
and generic primary fields, labeled by a continuous parameter $\alpha$, $\Phi_\alpha$ with dimension
\be
\Delta^M_\alpha=\alpha(\alpha-q).
\ee

\item \textbf{Ghost sector} needed to gauge fix the conformal symmetry. This is described by a fermionic $bc$ system of spin $(2,-1)$
of central charge
\be
c_\text{gh}=-26.
\ee

\end{enumerate}
The fact that
\be\label{c26}
c_L+c_M=26
\ee
allows the construction of a BRST complex.

The vertex operators of the complete system are built out of primary operators in the Liouville plus matter
sector as   
\be
U_\alpha=\Phi_{\alpha-b} V_{\alpha}
\ee
which are $(1,1)$-forms with ghost number zero, and can be integrated on the space.
This is ensured by the condition
\be
\Delta^M_{\alpha-b}+\Delta^L_{\alpha}=1.
\ee
We are mainly interested in three-point numbers and conformal blocks.
The former have been computed in \cite{Zamolodchikov:2005fy} (eq.7.9) for three generic dressed operators $U_{\alpha_i}$
and can be written in terms of $\gamma$ function, see eq. (\ref{gamma}), as 
\be\label{3pointLG}
\begin{aligned}
&C^\text{LG}(\alpha_1,\alpha_2,\alpha_3)
            =C^L(\alpha_1,\alpha_2,\alpha_3)C^M(\alpha_1-b,\alpha_2-b,\alpha_3-b) \\
            &=\left(\pi\mu\gamma(b^2)\right)^{(Q-\sum_{i=1}^3\alpha_i)/b}
            \left[\frac{\gamma(b^2)\gamma(b^{-2}-1)}{b^2}\right]^{1/2}
            \prod_{i=1}^3 \left[\gamma(2\alpha_i b-b^2)\gamma(2\alpha_i b^{-1}-b^{-2})\right]^{1/2}.
\end{aligned}
\ee

Let us remark that the ghost sector does not play any role in our considerations.
Indeed this is suited to produce a proper measure on the moduli space of the Riemann surface over which the CFT is formulated.
On the gauge theory side this would correspond to the quite unnatural operation of integrating over the gauge coupling.

\subsection{LG three-point function versus one-loop in gauge theory}

Let us now compare the results of the one-loop gauge theory partition function with the above three-point number of LG.
We consider the sector with zero magnetic fluxes $\vec m=\vec n=0$ of $U(2)$ gauge theory with $N_f=4$. 
The contribution of the one-loop partition function is given by equation (\ref{Z1loop-flux0}), and setting
$a_{12}=:2a$, we have
\be\label{Z1loop-flux0bis}
Z^{S^2\times S^2}_{\text{1-loop}}(\vec a,\ve_1,\ve_2)=\prod_{\alpha\neq\beta}a_{\alpha\beta}
=\prod_{\pm}\pm 2a
=-4a^2.
\ee
Indeed one can show that in the sector  $\vec m=\vec n=0$ the contribution of hypermultiplets in the four patches  (\ref{resHyp}) cancel each other.

The above result can be compared with the product of Liouville gravity three point numbers (\ref{3pointLG}).
Indeed, if we consider
\be\label{alpha_i}
\alpha=\frac{Q}{2}+a, \qquad \alpha_i=\frac{Q}{2}+p_i, \qquad a,p_i\in i\mathbb{R}
\ee
we get for the product of two three point numbers
\be\label{2-3pointsLG}
C^\text{LG}(\alpha_1,\alpha_2,\alpha)C^\text{LG}(\bar\alpha,\alpha_3,\alpha_4)
=N\left(\prod_{i=1}^4 f(\alpha_i)\right)\left(4a^2\right),
\ee
where
\be\label{vertices-renorm}
\begin{aligned}
N&=\left(\pi\mu\gamma(b^2)\right)^{1+b^{-2}}\frac{\gamma(b^2)\gamma(b^{-2}-1)}{b^2}, \\
f(\alpha_i)&=\left(\pi\mu\gamma(b^2)\right)^{-\alpha_i/b}\sqrt{\gamma(2\alpha_i b-b^2)\gamma(2\alpha_i b^{-1}-b^{-2})}.
\end{aligned}
\ee

The dependence on $a$ of (\ref{Z1loop-flux0bis}) and (\ref{2-3pointsLG}) is the same.
Moreover one can check \cite{Alday:2009aq} that the contribution of the two patches with $\ve_1^\li\cdot\ve_2^\li>0$
naturally compares to the product of Liouville theory three-point numbers (\cite{Zamolodchikov:2005fy}, eq. (2.2)).
On the other hand, the contribution of the two patches with $\ve_1^\li\cdot\ve_2^\li<0$
naturally compares to the product of generalized minimal model three-points functions
(\cite{Zamolodchikov:2005fy}, eq. (5.1) with $\beta=b$ and $\alpha=a-b$),
which is the matter sector of the LG.
Explicitly
\be
\begin{aligned}
&\left|Z^{S^2\times S^2}_{\text{1-loop}}(\vec a,\ve_1^\li\cdot\ve_2^\li>0)\right|
 =C^L(\alpha_1,\alpha_2,\alpha)C^L(\bar\alpha,\alpha_3,\alpha_4), \\[0.4cm]
&\left|Z^{S^2\times S^2}_{\text{1-loop}}(\vec a,\ve_1^\li\cdot\ve_2^\li<0)\right|
 =C^M(\alpha_1-b,\alpha_2-b,\alpha-b)C^M(\bar\alpha-b,\alpha_3-b,\alpha_4-b),
\end{aligned}
\ee
up to renormalization of the vertices analogously to (\ref{vertices-renorm}),
once the v.e.v. $a$ and $\mu_f$ are assumed to be purely imaginary.

We expect the gauge theory sectors with non-vanishing magnetic fluxes 
$\vec m$ and $\vec n$ to be related to the insertions of degenerate fields.
Indeed the same comment applies to the results on the conformal blocks obtained 
in the next subsection.

\subsection{Conformal blocks versus instantons}

It is a well known fact that the instanton contribution to the partition function
(\ref{ZinstC2matter}) for $U(2)$ gauge theory with $N_f=4$ on $\mathbb{C}^2$
can be matched with the four point conformal block on the sphere,
up to a $U(1)$ factor \cite{Alday:2009aq},
\be
Z^{\mathbb{C}^2,U(2)}_{\text{instanton}}(\epsilon_1,\epsilon_2, \vec{a}, \mu_1,\mu_2,\mu_3,\mu_4 )
\simeq
{\mathcal{F}^L}_{\alpha_1\phantom{\alpha_2}\alpha\phantom{\alpha_3}\alpha_4}^{
\phantom{\alpha_1}\alpha_2\phantom{\alpha}\alpha_3}(\tau)
\ee
where $\mu_1=p_1+p_2$, $\mu_2=p_1-p_2$, $\mu_3=p_3+p_4$, $\mu_4=p_3-p_4$
and $\alpha_i$ are defined in (\ref{alpha_i}).

Moreover 
contrary to three-point correlators,
the conformal blocks of the matter sector in LG
are the analytic continuation of those of Liouville theory under $b\to ib$.
These two facts allow us to interpret the instanton partition function (\ref{ZinstF0})
of $U(2)$ gauge theory with $N_f=4$ on $S^2\times S^2$ in the sector $\vec m=\vec n=0$
as two copies of four point conformal blocks of LG on the sphere.
Indeed, by using $b^\li=\sqrt{\ve_1^\li/\ve_2^\li}$, we have from (\ref{ZinstF0}) and (\ref{weights})
\be\label{ZinstF0}
\begin{aligned}
&Z^{S^2\times S^2}_{\text{instanton}}(b,b^{-1}, a, \mu_f ) \\
&=\prod_{\ell=1}^{4}
Z^{\mathbb{C}^2}_{\text{instanton}}(b^\li,(b^\li)^{-1}, a, \mu_f ) \\
&=\left[
Z^{\mathbb{C}^2}_{\text{instanton}}(b,b^{-1}, a, \mu_f )
  Z^{\mathbb{C}^2}_{\text{instanton}}(ib,(ib)^{-1}, a, \mu_f )\right]^2
\end{aligned}
\ee
where $\mu_f=\{\mu_1,\mu_2,\mu_3,\mu_4\}$.
From the discussion above, these are two copies of four points conformal blocks
of the two sectors of LG: Liouville and matter
\be
Z^{S^2\times S^2}_{\text{instanton}}(q,b,b^{-1}, a, \mu_f )\simeq
\left[
{\mathcal{F}^L}_{\alpha_1\phantom{\alpha_2}\alpha\phantom{\alpha_3}\alpha_4}^{
                 \phantom{\alpha_1}\alpha_2\phantom{\alpha}\alpha_3}(\tau)
{\mathcal{F}^M}_{\alpha_1\phantom{\alpha_2}\alpha\phantom{\alpha_3}\alpha_4}^{
                 \phantom{\alpha_1}\alpha_2\phantom{\alpha}\alpha_3}(\tau)
\right]^2=
\left[
{\mathcal{F}^\text{LG}}_{\alpha_1\phantom{\alpha_2}\alpha\phantom{\alpha_3}\alpha_4}^{
                          \phantom{\alpha_1}\alpha_2\phantom{\alpha}\alpha_3}(\tau)
\right]^2.
\ee
The full partition function (\ref{Zfull}) in the sector $\vec m=\vec n=0$ is then expressible as
\be
Z^{S^2\times S^2}_{\text{full}}(q,b,b^{-1}, a, \mu_f )
\propto \int d\alpha\, C^\text{LG}(\alpha_1,\alpha_2,\alpha)C^\text{LG}(\bar\alpha,\alpha_3,\alpha_4)
           \left[{\mathcal{F}^\text{LG}}_{\alpha_1\phantom{\alpha_2}\alpha\phantom{\alpha_3}\alpha_4}^{
                          \phantom{\alpha_1}\alpha_2\phantom{\alpha}\alpha_3}(\tau)\right]^2.
\ee
Few remarks are in order here. First of all, the holomorphicity in its arguments of the supersymmetric partition function under scrutiny is reflected in the holomorphic gluing of building blocks of the corresponding CFT,
in contrast to the one appearing in correlation functions of Liouville gravity. Moreover,
we underline that the three-point numbers of the matter sector $C^M$ naturally arising in the gauge theory context
are strictly speaking not the ones of generalised minimal model. Indeed they do not obey the selections rules
of this model, and were introduced in \cite{Zamolodchikov:2005fy} only as a technical tool to solve the relevant
bootstrap equations. Indeed to get the physical three-point functions one has to multiply them by a suitable non-analytic term
which takes into account the selection rule (see eq(3.16) in \cite{Zamolodchikov:2005fy} and also \cite{0510214}). We remark that a CFT which
consistently makes use of the analytic three-point correlator $C^M$ appearing in the gauge theory can be formulated
\cite{RS}.

\section{Discussion}
\label{sec:conclusion}

In this paper we computed the partition function of $\N=2$ supersymmetric gauge theory on $S^2\times S^2$.
We derived the generalized Killing spinor equations for the extended supersymmetry to exist on a four manifold by consistency of the supersymmetry algebra and found 
a slight generalization of the ones discussed in \cite{Hama:2012bg}.
We derived spinor solutions to these equations realizing a version of Witten's topological twist which is equivariant with respect to
a $U(1)$ isometry of the manifold, and then exploited them to construct a supercharge localizing on the fixed points
of the isometry.
The resulting partition function is defined by gluing Nekrasov partition functions and integrating the v.e.v. of the scalar field $\Phi$
of the twisted vector multiplet over a suitable contour.
 We also showed that the resulting partition function displays the three-point correlators and conformal blocks of
Liouville gravity as building blocks. Notice however that these are glued in a different way with respect to Liouville gravity correlators.
In particular our partition function is holomorphic in the momenta of the vertices and in their positions. It would be interesting
to investigate further if there is a chiral conformal field theory interpretation of the gauge theory result.
Notice that chiral correlation functions for Liouville theory can be defined for some special values of the central charge
by using the relation with super Liouville suggested by gauge theory \cite{Bonelli:2011jx,Bonelli:2011kv} and further investigated in \cite{Belavin:2011sw,Schomerus:2012se,Hadasz:2013dza}.
In this case the field theory is defined over the resolution of $\mathbb{C}^2/\mathbb{Z}_2$, whose projective compactification
is the second Hirzebruch surface $\mathbb{F}_2$. This is a framework very near to the one investigated in the present paper.
Indeed, let us stress once again that although we focused on $\mathbb{P}^1\times \mathbb{P}^1$, which coincides with the Hirzebruch surface $\mathbb{F}_0$,
our approach can be easily extended to a wide class of manifolds, including compact toric ones. 
In this context, it is conceivable that our results can be used to prove a long standing conjecture by Nekrasov proposing
a contour integral formula for Donaldson's invariants and its generalization to $SU(N)$ gauge groups.

Let us underline that in the gauge theory there are two consistent choices of reality conditions for the fields $\Phi,\chir \Phi$  \cite{Witten:1988ze},
either real and independent or $\chir \Phi= - \Phi^\dagger$. This leads to the choice of different 
integration contours which it would be interesting to investigate in the conformal field theory counterpart.

Another possible check of the relation of the $\N=2$ gauge theory on  $S^2\times S^2$ and Liouville gravity would be to compute the 
$\N=2^*$ case and compare with the CFT on the one-punctured torus.

We also discussed another solution to the Killing spinor equation on  $S^2\times S^2$ (see Appendix C)
which are composed by the spinorial solutions on $S^2$ discussed in \cite{Benini:2012ui,Doroud:2012xw}.
These solutions correspond to trivial ${\cal R}$-symmetry bundle and therefore imply the Witten topological twist only locally, more precisely
they localize on instantons on the $NN$ and $SS$ fixed points of $S^2\times S^2$ and on anti-instantons on $NS$ and $SN$.
The partition function in this case is neither real nor holomorphic and 
it would be interesting to further discuss the supersymmetric path integral induced by these solutions.

Another related subject to investigate is the reduction to spherical partition functions in the zero volume limit of one of the two 
spheres. This analysis would help in shedding light on the relation between instanton and vortex partition functions.
It would be interesting to consider the insertion of surface operators 
\cite{Alday:2009fs,Drukker:2009id}
on one of the two spheres.
Actually, surface operators on $\mathbb{C}^2$ are related to the moduli space of instantons on $S^2\times S^2$ framed on one of the 
two spheres \cite{Awata:2010bz}
so that a nice interplay could arise among these partition functions.

As already stated, our equivariant localization scheme applies to much general cases than the one on which we focus here. 
It would be very interesting to extend our approach to other four manifolds and in this framework analyze possible relations between 
Gromov-Witten and Donaldson invariants. This could open further applications of the gauge theory computations to integrable systems.

It would be very interesting also to study the holographic dual of the large $N$ limit of the partition function 
on $S^2\times S^2$ that we just computed.

\section*{Acknowledgments}

We thank
A.~Belavin,
V.~Belavin, 
M.~Bershtein, 
S.~Cremonesi,
K.~Hosomichi,
B.~Mares,
K.~Maruyoshi,
R.~Santachiara
and
A.~Sciarappa
for useful discussions.

This research was partly supported 
by the INFN Research Projects GAST and ST\&FI
and 
by PRIN ``Geometria delle variet\`a algebriche".

\appendix

\section{Full ${\cal N}=2$ Supersymmetry}
\label{N=2}

In this appendix we give the detailed calculations of the results stated in subsection (\ref{full}).

We proceed by writing the most general form for supersymmetric variation for the gauginos, consistent with the properties of positivity of mass of background fields,
their gauge neutrality, and balancing of masses and $U(1)_R$ charges. We have
\be
\begin{aligned}
 \susy\lambda_A &=
 \frac{k_1}{2}\sigma^{\m\n}\xi_A(F_{\m\n}+8\bar\phi T_{\m\n} + 8\phi W_{\m\n})
 +a_1\sigma^\m\bar\xi_AD_\m\phi
 +b_1\sigma^\m D_\m\bar\xi_A\phi+c_1\xi_A[\phi,\bar\phi]
 +d_1 D_{AB}\xi^B,
 \\
 \susy\bar\lambda_A &=
 \frac{k_2}{2}\bar\sigma^{\m\n}\bar\xi_A(F_{\m\n}+8\phi\bar T_{\m\n} + 8\bar\phi \bar W_{\m\n})
 +a_2\bar\sigma^\m\xi_AD_\m\bar\phi
 +b_2\bar\sigma^\m D_\m\xi_A\bar\phi-c_2\bar\xi_A[\phi,\bar\phi]+d_2 D_{AB}\bar\xi^B.
 \\
\end{aligned}
\ee

Consider now the square of the supersymmetry transformation acting on the scalar fields
\be
\begin{aligned}
 \susy^2\phi &= -i\xi^A(\susy\lambda_A) = -ia_1\xi^A\sigma^\m\bar\xi_AD_\m\phi -i b_1\xi^A\sigma^\m D_\m\bar\xi_A\phi - ic_1\xi^A\xi_A[\phi,\bar\phi], \\
 \susy^2\bar\phi &= -i\bar\xi^A(\susy\bar\lambda_A) = +ia_2\bar\xi^A\bar\sigma^\m\xi_AD_\m\bar\phi +i b_2\bar\xi^A\bar\sigma^\m D_\m\xi_A\bar\phi 
- ic_2\bar\xi^A\bar\xi_A[\phi,\bar\phi].
\end{aligned}
\ee
From the above, it clearly follows that $V^\m = ia_1\bar\xi^A\bar\sigma^\m\xi_A = ia_2\bar\xi^A\bar\sigma^\m\xi_A$. We define $a \equiv a_1 = a_2$. We also infer that 
$\Phi = c_1 \xi^A\xi_A\bar\phi - c_2 \bar\xi^A\bar\xi_A\phi$. We will return to $\susy^2\phi$ and $\susy^2\bar\phi$ momentarily to investigate the scaling and the $U(1)_R$ terms.

Consider now the  $\susy^2A_\m$
\be
\begin{aligned}
 \susy^2A_\m =& i\xi^A\sigma_\m(\susy\bar\lambda_A) - i\bar\xi^A\bar\sigma_\m(\susy\lambda_A) \\
=& \frac{i}{2}F^{\rho\sigma}\left( k_2\xi^A\sigma_\m\bar\sigma_{\rho\sigma}\bar\xi_A -k_1 \bar\xi^A\bar\sigma_\m\sigma_{\rho\sigma}\xi_A \right) \\
&+ 4i\left( k_2\phi \bar T^{\rho\sigma}\xi^A\sigma_\m\bar\sigma_{\rho\sigma}\bar\xi_A
  -k_1\bar \phi T^{\rho\sigma}\bar\xi^A\bar\sigma_\m\sigma_{\rho\sigma}\xi_A \right)\\
&+ 4i\left( k_2\bar\phi \bar W^{\rho\sigma}\xi^A\sigma_\m\bar\sigma_{\rho\sigma}\bar\xi_A
  -k_1\phi W^{\rho\sigma}\bar\xi^A\bar\sigma_\m\sigma_{\rho\sigma}\xi_A \right)\\
&+ ia \left( \xi^A \sigma_\m\bar\sigma_\n\xi_A D^\n \bar\phi - \bar\xi^A \bar\sigma_\m \sigma_\n\bar\xi_A D^\n \phi \right)\\
&+ i\left(b_2  \bar\phi \xi^A \sigma_\m\bar\sigma_\n  D^\n \xi_A - b_1  \phi \bar\xi^A \bar\sigma_\m\sigma_\n  D^\n \bar\xi_A \right) \\
&- i[\phi,\bar\phi] \left(c_1 \xi^A\sigma_\m \bar\xi_A + c_2 \bar\xi^A \bar\sigma_\m\xi_A \right) \\
&- iD_{AB} \left( d_2 \xi^A \sigma_\m \bar\xi^B - d_1 \bar\xi^A \bar\sigma_\m \xi^B \right)
\end{aligned}
\ee
The commutator term must vanish by the assumptions on the nature of $\susy^2$, which implies  $c_1 = c_2 \equiv c$. Similarly the vanishing of the $D_{AB}$ requires $d_1 = d_2$, 
which can now be absorbed in $D_{AB}$, and we will therefore set $d_1 = d_2 =1$.
We want the term with $F^{\rho\sigma}$ to equal $V^\n F_{\n\m}$, which forces $k_1 = k_2 = (a/2) \equiv k$ as can be seen after some algebraic manipulations of the spinor products.
The terms that remain are the ones with the background fields and the ones with the derivatives of the scalar field:
\begin{equation}
\begin{aligned}
 &4i\left( k_2\phi \bar T^{\rho\sigma}\xi^A\sigma_\m\bar\sigma_{\rho\sigma}\bar\xi_A
  -k_1\bar \phi T^{\rho\sigma}\bar\xi^A\bar\sigma_\m\sigma_{\rho\sigma}\xi_A \right)\\
+&4i\left( k_2\bar\phi \bar W^{\rho\sigma}\xi^A\sigma_\m\bar\sigma_{\rho\sigma}\bar\xi_A
  -k_1\phi W^{\rho\sigma}\bar\xi^A\bar\sigma_\m\sigma_{\rho\sigma}\xi_A \right)\\
+&ai \left( \xi^A \sigma_\m\bar\sigma_\n\xi_A D^\n \bar\phi - \bar\xi^A \bar\sigma_\m \sigma_\n\bar\xi_A D^\n \phi \right).
\end{aligned}
\end{equation}
We require these terms to be equal to the gauge variation
\be
 D_\m\Phi =  c \xi^A\xi_A D_\m \bar\phi - c \bar\xi^A\bar\xi_AD_\m \phi
          + 2c\bar\phi\xi^AD_\m\xi_A - 2c\phi\bar\xi^AD_\m\bar\xi_A.
\ee
Equating the terms with the derivatives of the scalar field on the two sides
gives $c= ia$ while equating the terms in $\bar\phi$ we get
\begin{equation}
2ia\xi^AD_\m\xi_A = -4ik (T^{\rho\sigma}-W^{\rho\sigma})\xi^A \sigma_{\rho\sigma}\sigma_\m \bar\xi_A
+ ib_2\xi^A\bar\sigma_\n D^\n\xi_A.
\end{equation}
We note that this is satisfied when
\begin{equation}
aD_\m\xi_A = -2k(T^{\rho\sigma}-W^{\rho\sigma})\sigma_{\rho\sigma}\sigma_\m \bar\xi_A 
+ \frac{b_2}{2}\sigma_\m \bar\sigma_\n D^\n\xi_A. \label{eq:tentativeKilling1}
\end{equation}
Contracting either side with $\sigma^\m$ we find that $a = 2b_2$. Similarly starting with the equation for 
$\phi$ we  find that $a = 2b_1$. We also find analogously the equation
\begin{equation}
aD_\m\bar\xi_A = -2k(\bar T^{\rho\sigma}-\bar W^{\rho\sigma})\bar\sigma_{\rho\sigma}\bar\sigma_\m \xi_A 
+ \frac{b_2}{2}\bar\sigma_\m \sigma_\n D^\n \bar\xi_A. \label{eq:tentativeKilling2}
\end{equation}
Define $b \equiv b_1 = b_2$.
We now return to expressions for $\susy^2\phi$,  $\susy^2\bar\phi$ and $\susy^2 A_\m$ and identify the remaining terms.
Consider first the equation for $\susy^2\phi$ and  $\susy^2\bar\phi$. Since we have identified the Lie derivative term and 
the gauge transformation term, the remaining terms must combine to give the scaling and the $U(1)_R$ terms.
Therefore
\begin{equation}
\begin{aligned}
4ik (D_\m \bar\xi^A\bar\sigma^\m\xi_A + \bar\xi^A\bar\sigma^\m D_\m\xi_A) + 2\Theta 
&= ib_1 D_\m \bar\xi^A\bar\sigma^\m\xi_A \\
4ik (D_\m \bar\xi^A\bar\sigma^\m\xi_A + \bar\xi^A\bar\sigma^\m D_\m\xi_A) - 2\Theta 
&= ib_2  \bar\xi^A\bar\sigma^\m D_\m\xi_A
\end{aligned}
\end{equation}
which gives
\begin{equation}
\Theta = \frac{ib}{4}(D_\m \bar\xi^A\bar\sigma^\m\xi_A - \bar\xi^A\bar\sigma^\m D_\m\xi_A).
\label{eq:Thetaformula}
\end{equation}
Consider now $\susy^2 A_\m$. We are left with the following terms.
\begin{equation}
\begin{aligned}
&4i\left( k_2\phi \bar T^{\rho\sigma}\xi^A\sigma_\m\bar\sigma_{\rho\sigma}\bar\xi_A
  -k_1\bar \phi T^{\rho\sigma}\bar\xi^A\bar\sigma_\m\sigma_{\rho\sigma}\xi_A \right)\\
&+ 4i\left( k_2\bar\phi \bar W^{\rho\sigma}\xi^A\sigma_\m\bar\sigma_{\rho\sigma}\bar\xi_A
  -k_1\phi W^{\rho\sigma}\bar\xi^A\bar\sigma_\m\sigma_{\rho\sigma}\xi_A \right)\\
&+ ia \left( \xi^A \sigma_\m\bar\sigma_\n\xi_A D^\n \bar\phi - \bar\xi^A \bar\sigma_\m \sigma_\n\bar\xi_A D^\n \phi \right).
\end{aligned}
\end{equation}
We require that these combine to give the appropriate gauge transformation term
\begin{equation}
D_\m\big[2ik(\xi^A\xi_A \phi - \bar\xi^A\bar\xi_A \bar\phi)\big]
\end{equation}
which happens when equations \eqref{eq:tentativeKilling1} and \eqref{eq:tentativeKilling2} are satisfied.

Note that we can rescale the gauginos and the auxiliary field to get rid to the normalization $k$ (or 
equivalently $a$, $b$ or $c$). We therefore set $k=1$. We summarize the expressions for the generators of the 
bosonic symmetries that we have found till now:
\begin{equation}
\begin{aligned}
V^\m &= 2i \bar\xi^A\bar\sigma^\m\xi_A, \\
w &=\frac{1}{4}D_\m v^\m \\
\Theta &= \frac{i}{4}(D_\m \bar\xi^A\bar\sigma^\m\xi_A - \bar\xi^A\bar\sigma^\m D_\m\xi_A) \\
\Phi &= 2i\bar\phi \xi^A\xi_A - 2i\phi\bar\xi^A\bar\xi_A.
\end{aligned}
\end{equation}

We now study $\susy^2 \lambda_A$. The case of $\susy^2 \bar\lambda_A$ is analogous and will not be detailed. 
In doing so, will find the expression for $\susy D_{AB}$ and also show that $W_{\m\n}$ vanishes.
\begin{equation}
\begin{aligned}
\susy^2\lambda_A =& \frac{1}{2}\sigma^{\m\n}\xi_A\left(\susy F_{\m\n} + 8(\susy \bar\phi)T_{\m\n}
+ 8(\susy \phi)W_{\m\n} \right)\\
&+ 2\sigma^\m \bar\xi_A\big(D_\m(\susy\phi) - i[\susy A_\m,\phi]\big) +  \sigma^\m D_\m\bar\xi_A (\susy \phi)\\
&+ 2i\xi_A[\susy\phi,\bar\phi] + 2i\xi_A[\phi.\susy\bar\phi] + \susy D_{AB}\xi^B \\
=& \frac{i}{2}\sigma^{\m\n}\xi_A \Big[ D_\m \xi^B\sigma_\n\bar\lambda_B + \xi^B\sigma_\n\bar D_\m \lambda_B \\
& -D_\m \bar\xi^B\bar\sigma_\n\lambda_B - \bar\xi^B\bar\sigma_\n D_\m \lambda_B\\
& -D_\n \xi^B\sigma_\m\bar\lambda_B - \xi^B\sigma_\m\bar D_\n \lambda_B\\
& +D_\n \bar\xi^B\bar\sigma_\m\lambda_B + \bar\xi^B\bar\sigma_\m D_\n \lambda_B\Big]\\
&+ 4i\sigma^{\m\n}\xi_A T_{\m\n}(\bar\xi^B\bar\lambda_B) -4i\sigma^{\m\n}\xi_A W_{\m\n}(\xi^B\lambda_B)\\
&- 2i\sigma^\m\bar\xi_A (D_\m\xi^B\lambda_B) - 2i\sigma^\m\bar\xi_A (\xi^B D_\m\lambda_B)\\
&+ 2\sigma^\m\bar\xi_A[\xi^B\sigma_\m \bar\lambda_B -\bar\xi^B\bar\sigma_\m \lambda_B, \phi ] -i\sigma^\m D_\m\bar\xi_A (\xi^B\lambda_B) \\
&-2\xi_A[\bar\phi,\xi^B\lambda_B]  -2\xi_A[\phi,\bar\xi^B\bar\lambda_B] \\
&+(\susy D_{AB})\xi^B.
\end{aligned}
\end{equation}
Since we know what form we should force $\susy^2 \lambda_A$ to take, we can rearrange the above terms to obtain them. 
We try to cancel all the offending terms by postulating the form for the supersymmetric variation of the auxiliary field (as we did in the chiral case). We find that 
\begin{equation}
\susy D_{AB} =-2i\chir\xi_{(A}\chir\s^\m D_\m\lambda_{B)}+2i\xi_{(A}\s^\m D_\m \chir\lambda_{B)}
                   -4\left[\phi,\chir\xi_{(A}\chir\lambda_{B)}\right]+4\left[\chir\phi,\xi_{(A}\lambda_{B)}\right].
\end{equation}
We also find the form of $\Theta_{AB}$ as follows:
\begin{equation}
\Theta_{AB} = -i\xi_{(A}\sigma^\m D_\m \bar\xi_{B)} + iD_\m \xi_{(A}\sigma^\m \bar\xi_{B)}. 
\end{equation}
We rediscover the main Killing equation as the co-efficient of the terms with $\bar\lambda$
\begin{equation}
2i(\bar\lambda^B\bar\sigma^\m\xi_A)\left( D_\m \xi_B + T^{\rho\sigma}\sigma_{\rho\sigma}\sigma_\m\bar\xi_B - \frac{1}{4} \sigma_{\m}\bar\sigma_{\n}D^{\n}\xi_B\right).
\end{equation}

And finally, consider the term with $W_{\m\n}$:
\begin{equation}
-4i\sigma_{\m\n}W^{\m\n}(\xi^B\lambda_B)\xi_A = -4i\sigma_{\m\n}W^{\m\n}\left[ (\xi^B\xi_B)\lambda_A
+ (\xi_A\lambda_B + \xi_B\lambda_A)\xi^B \right].
\end{equation}
We see that while the second parenthesis can possibly be absorbed into $\susy  D_{AB}$, the first term remains and does not fit the desired form for $\susy^2 \lambda_A$ which implies that $W_{\m\n} = 0$. Finally, let us consider $\susy^2 D_{AB}$. After a routine calculation, and using the main Killing equation we recover the expression written above, except one failure term:
\begin{equation}
\begin{aligned}
&-i\phi(\bar\xi_{(A}\bar\sigma^\m\sigma^\n D_\m D_\n\bar\xi_{B)} - 4\xi_{(A}\sigma^\m\bar\sigma^{\rho\sigma}\bar\xi_{B)} D_{\m}\bar T_{\rho\sigma}) \\ \nonumber
& +i \bar\phi(\xi_{(A}\sigma^\m \bar\sigma^\n D_\m D_\n\xi_{B)} - 4\bar\xi_{(A}\bar\sigma^\m\sigma^{\rho\sigma}\bar\xi_{B)} D_{\m}T_{\rho\sigma})
\end{aligned}
\end{equation}
which yield the following two auxiliary equations
\begin{equation}
\begin{aligned}
\s^\m \chir\s^\n D_\m D_\n \xi_A + 4 D_\lambda T_{\m\n} \s^{\m\n} \s^\lambda \chir\xi_A &= M_1\xi_A, \\
\chir \s^\m \s^\n D_\m D_\n \chir\xi_A + 4 D_\lambda \chir T_{\m\n} \chir\s^{\m\n} \chir\s^\lambda \xi_A &= M_2\chir\xi_A.
\end{aligned}
\end{equation}
We note that there is no reason, at the level of the supersymmetry algebra from the above approach, to have a single scalar background field $M$.

\section{Generic twisting solutions}
\label{app:GenSol}
It is possible to derive solutions for (\ref{eq:1bis}) (\ref{eq:2}) of the same kind of (\ref{eq:ResumeSol})
for a generic four-manifold admitting a $U(1)$ isometry generated by a Killing vector $V$.

Such a solution will generate the vector $V$ as in (\ref{eq:KillVec}).

In this general setting we have to turn on the whole $SU(2)_R$ bundle
and the Witten twist (\ref{eq:Wittentwist}) becomes
\be
G_{\m\, A}^{\phantom{\m A}B}=\sum_{k=1}^3 G^k_\mu\, \s^{(k)B}_{\;\,A}
\ee
with
\be
G^1=-\frac{1}{2}(\omega^{14}+\omega^{23}), \quad
G^2=-\frac{1}{2}(\omega^{13}-\omega^{24}), \quad
G^3=-\frac{1}{2}(\omega^{12}+\omega^{34}),
\ee
where $\omega^{ab}$ denote the components of the spin connection one-form.
This twist admits the following solution for equations (\ref{eq:1bis}) (\ref{eq:2})
\be
\xi_1=\frac{i}{4}\left( \begin{array}{c}  V_3+iV_4 \\  V_1+iV_2 \end{array} \right), \quad
\xi_2=\frac{i}{4}\left( \begin{array}{c}  V_1-iV_2 \\ -V_3+iV_4 \end{array} \right), \quad
\chir\xi_1=\left( \begin{array}{c} 1 \\ 0 \end{array} \right), \quad
\chir\xi_2=\left( \begin{array}{c}  0 \\ 1 \end{array} \right),
\ee
(where $V_a=e_a^\mu V_\mu$) and the background fields are chosen as
\be
T=-\frac{1}{32}(d\zeta)^-, \quad \chir S=\frac{1}{32}(d\zeta)^+, \quad \chir T=0, \quad S=0,  \quad M=0,
\ee
where the superscripts $-$ and $+$ denote the anti self-dual and the self-dual part respectively
and $\zeta=\star\iota_V \star 1$.

\section{Untwisted solutions}
\label{untwisted-sol}

In this appendix we summarize solutions to equations (\ref{eq:1}),(\ref{eq:2}) that follow from assumptions of the vanishing of $SU(2)_R$ gauge field
and the direct product decomposition of $\xi_A$ in terms of conformal Killing spinors on 2-spheres. However, these solutions have 
the disadvantage of not being real in the sense of equation (\ref{eq:reality}). We will therefore not be using the solutions derived in
this appendix in the rest of the paper, but will summarize them here for possible future applications.

To solve the equations we consider the following ansatze:
\begin{itemize}
\item The $SU(2)_R$ gauge field is zero
      \be
      {G_\m}^B_{\,A} = 0.
      \ee
      Since this condition implies there is no mixing between different $SU(2)_R$ components,
      we will drop the indices $A,B,\dots$ for the remainder of this subsection.

\item The background fields $T_{\m\n}$, $\chir T_{\m\n} $ are respectively anti self-dual and self-dual combination
of the two-dimensional volume forms in the two sphere $\omega\1,\,\omega\2$.
\be
\begin{aligned}
T\equiv \frac{1}{2} T_{\m\n} \, dx^\m \wedge dx^\n &= t(\omega\1-\omega\2), \\
\chir T\equiv \frac{1}{2} \chir T_{\m\n} \, dx^\m \wedge dx^\n &= \chir t (\omega\1+\omega\2).
\end{aligned}
\ee
where $t$ and $\chir t$ are complex numbers (the bar does not imply that they are complex conjugates).

\item The candidate solution $\xi$ is a tensor product of two-dimensional Killing spinors on each sphere
\be
\xi=\ve\1 \otimes \ve\2
\ee
where $\ve\1=\ve\1(\theta_1,\varphi_1)$ and $\ve\2=\ve\2(\theta_2,\varphi_2)$.
The spinor on the right hand side, $\xi'$ has an analogous decomposition $\xi'=\ve'\1 \otimes \ve'\2$.

\end{itemize}
Through these assumptions, we intend to decompose (\ref{eq:1}) and (\ref{eq:2})
into tensor products of equations on either spheres.
To do this we use the following representation for the gamma matrices
\begin{equation} \label{eq:reprtens}
\begin{array}{lll}
\g_1 = \s_1 \otimes \Id, \quad & \g_2 = \s_2 \otimes \Id, & \phantom{a} \\
\g_3 = \s_3 \otimes \s_1, \quad & \g_4 = \s_3 \otimes \s_2, &
\quad \g_5  = \s_3\otimes\s_3 =-\g_1\g_2\g_3\g_4.
\end{array}
\end{equation}
Using these facts and the ansatze at the beginning of this section for the background fields and spinors we obtain
\be \label{eq:split}
\begin{aligned}
& (D\1 \ve\1 \otimes \ve\2) + (\ve\1 \otimes D\2 \ve\2)  \\
& \!\!+ i(\chir t + t) \big[ (\s_3 \, \s\1 \ve\1 \otimes \ve\2) + ( \ve\1 \otimes \s\2 \ve\2 ) \big]  \\
& \!\!+ i(\chir t - t) \big[ (\s\1 \ve\1 \otimes \s_3 \, \ve\2) + (\s_3 \, \ve\1 \otimes \s_3 \, \s\2 \ve\2) \big] \\
& \quad = -i(\s\1 {\ve'}\1 \otimes {\ve'}\2) -i (\s_3 \, {\ve'}\1 \otimes \s\2 {\ve'}\2)
\end{aligned}
\ee
where the labels {\footnotesize (1)} and {\footnotesize (2)} mean ``relative of the first and the second sphere'' respectively, and\footnote{
In this subsection we use the symbol $\Omega$ for the spin connection,
this is to avoid confusion with the volume forms.}
\be
\begin{aligned}
&D\1=d\1 +\frac{1}{2} \Omega\1^{12}\s_{12} &\qquad &    D\2=d\2 +\frac{1}{2} \Omega\2^{34}\s_{12} \\
&d\1=\de_{\theta_1} \, d\theta_1 + \de_{\varphi_1} \, d\varphi_1   &\qquad
&d\2=\de_{\theta_2} \, d\theta_2 + \de_{\varphi_2} \, d\varphi_2 \\
&\s\1=e^1 \s_1 + e^2 \s_2   &\qquad & \s\2=e^3 \s_1 + e^4 \s_2     \\
&\Omega\1^{ab}=\Omega_{\varphi_1}^{ab}\, d\varphi_1        \quad a,b=1,2   &\qquad
&\Omega\2^{a-2,b-2}=\Omega_{\varphi_2}^{ab}\, d\varphi_2   \quad a,b=3,4.
\end{aligned}
\ee
The vielbein are
\be
       e^1=r_1 d\theta_1 \quad  e^2=r_1 \sin\theta_1 d\varphi_1 \quad 
       e^3=r_2 d\theta_2 \quad  e^4=r_2 \sin\theta_2 d\varphi_2.
\ee

The conformal Killing spinors in two dimension for the $S^2$ metric are already known \cite{Benini:2012ui}. These are spanned by the solutions to the following equations
\begin{equation}
D \ve_\pm = \pm \frac{i}{2r} e^a \s_a \ve_\pm \label{eq:BCsol1}.
\end{equation}
One can find an alternate basis for the Killing spinors, where the elements of the basis satisfy
\begin{equation}
D \hat\ve_\pm = \pm \frac{1}{2r} e^a \s_a \s_3 \hat\ve_\pm   \label{eq:BCsol2}.
\end{equation}
The two basis are related by
\be
\hat\ve_\pm = (\Id + i \s_3) \ve_\pm. \label{eq:basisrelation}
\ee
Corresponding to each sign, in either of the two equations, there are two linearly independent solutions.
For example, the solutions to equation (\ref{eq:BCsol1}) are given 
(up to normalization) by
\be
\begin{array}{ll}
\ve^{+,1}  = e^{-(i/2) \varphi}\left( \begin{array}{r} \sin\theta/2 \\ \!\! -i\cos\theta/2 \end{array} \right), & 
\ve^{+,2}  = e^{(i/2) \varphi}\left( \begin{array}{r}  \cos\theta/2 \\       i\sin\theta/2 \end{array} \right),  \\[4mm]
\ve^{-,1}  = e^{-(i/2)\varphi}\left( \begin{array}{r} \sin\theta/2 \\       i\cos\theta/2 \end{array} \right), &
\ve^{-,2}  = e^{(i/2) \varphi}\left( \begin{array}{r}  \cos\theta/2 \\ \!\! -i\sin\theta/2 \end{array} \right).
\end{array}
\ee
The linearly independent solutions to (\ref{eq:BCsol2}) may be found using the above solutions and 
equation (\ref{eq:basisrelation}).
It must be noted that the sign of $\ve$ does not indicate its chirality,
and indeed solutions of definite ``positivity'' do not have definite chirality.

We use the existence of these solutions to rewrite (\ref{eq:split}) into an algebraic equation.
To do so, let us notice that on the left hand side of equation (\ref{eq:split}),
we have the terms $(D\1 \ve\1 \otimes \ve\2)$ as well as of the form $(\s_3 \, \s\1 \ve\1 \otimes \ve\2)$. 
This suggests that we should take $\ve\1$ to be a solution of equation (\ref{eq:BCsol2}), and ${\ve'}\1$ to be proportional to $\s_3\ve\1$ (with $r = r_1$). Similarly, because the left hand side
of (\ref{eq:split}) contains $\ve\1 \otimes D\2 \ve\2$ and $\ve\1 \otimes \s\2 \ve\2 $, we are compelled to choose $\ve\2$ as a solution of (\ref{eq:BCsol1}) (with $r = r_2$) and 
${\ve'}\2$ to be proportional to $\ve\2$. Equation (\ref{eq:split}) then decomposes into two algebraic equations for Killing spinors on either spheres if we take the coefficient
$(\chir t - t) =0$ (as the terms with this coefficient do not conform to the pattern of the other terms
and to equations (\ref{eq:BCsol1}) and (\ref{eq:BCsol2})).

Incorporating these observations in (\ref{eq:split}), we get:

\begin{equation}
\begin{aligned}
 & (D\1 \ve\1 \otimes \ve\2) + (\ve\1 \otimes D\2 \ve\2)  \\
& + 2it \big[ (\s_3 \, \s\1 \ve\1 \otimes \ve\2) + ( \ve\1 \otimes \s\2 \ve\2 ) \big]  \\
& = -iC\big[(\s\1 \s_3\ve\1 \otimes {\ve}\2) -i (\ve\1 \otimes \s\2 {\ve}\2)\big]
\end{aligned}
\end{equation}
where $C$ is a proportionality constant; $\ve\1$ and $\ve\2$ are solutions of (\ref{eq:BCsol2}) and (\ref{eq:BCsol1}) respectively. It is obvious that
up on using equations (\ref{eq:BCsol2}) and (\ref{eq:BCsol1}) we are left with purely algebraic equations that can be easily solved for $t$ and $C$ in terms of $r_1$ and $r_2$.
We have four families of solutions in all, corresponding to four choices of signs that can be made. The solutions can be summarized as:
\begin{equation}
\xi = {\hat\ve}_{(1)}^\pm \otimes {\ve}_{(2)}^{\pm^\prime};\hspace{1cm} C= \frac{1}{4}\left( \pm \frac{i}{r_1} \mp^\prime \frac{1}{r_2} \right);\hspace{1cm}
 t=\chir t = \frac{1}{16} \left(\mp \frac{i}{r_1} \pm^\prime \frac{1}{r_2} \right).
\end{equation}

The auxiliary equation (\ref{eq:2}) may be decomposed in a similar manner. 
It turns out that the value of the scalar background field $M$ is the same for all four families of solutions and is given by
\be
M=-\left( \frac{1}{r_1^2}+\frac{1}{r_2^2} \right).
\ee

Note that since for each choice of sign in either equations (\ref{eq:BCsol2}) or (\ref{eq:BCsol1}), we have a 2-complex dimensional family of solutions, each family of solutions 
is the complex span of four linearly independent spinors.

Finally we would like to return to the standard Clifford algebra representation
\be\label{eq:reprstand}
\gamma_a=\left(
         \begin{array}{cc}
         0          &  \s_a \\
         \chir\s_a  &  0
         \end{array}
         \right)
\qquad a=1,\dots,4.
\ee
To do that we use the following unitary transformation
\be
T=\left( \begin{array}{cccc}
               0 & 1 & 0 & 0 \\
               0 & 0 & 1 & 0 \\
              i & 0 & 0 & 0 \\
               0 & 0 & 0 & i
       \end{array} \right)
\ee
We may then use two of the four solutions in any one given family, and put them in an $SU(2)_R$ doublet.

\section{Conventions}

\subsection{Notation}

Latin indices $\{a,b,\dots\}$ are used for flat space coordinates,
and are used for both real coordinates $a,b=1,2,3,4$ and complex coordinates $a,b=1,\bar 1,2,\bar 2$.
Greek indices $\{\m,\n,\dots\}$ are used for curved space coordinates,
real $\m,\n=\theta_1,\varphi_1,\theta_2,\varphi_2$ or complex $\m,\n=z,\bar z, w ,\bar w$.
Any ambiguities in this notation should be clarified from the context.

The metric in the flat space $\delta_{ab}$ is link with the metric in curved space $g_{\m\n}$
via the vierbein $e_\m^a$
\be
g_{\m\n}=e_\m^a e_\n^b \delta_{ab}.
\ee

\subsection{Metrics}
\label{App:metric}

The metric of $S^2\times S^2$ in real coordinates is
\be\label{eq:realmetric}
\begin{aligned}
ds^2&=g_{\m\n}dx^\m dx^\n=\delta_{ab} e^a e^b \\
    &=r_1^2(d\theta_1^2+\sin^2\theta_1 d\varphi_1^2)+r_2^2({d\theta_2}^2+\sin^2\theta_2 {d\varphi_2}^2).
\end{aligned}
\ee
Therefore the vierbein 1-forms $e^a=e^a_\m dx^\m$ are
\be
e^1=r_1 d\theta_1, \quad e^2=r_1\sin\theta_1 d\varphi_1, \quad e^3=r_2 d\theta_2, \quad e^4=r_2\sin\theta_2 d\varphi_2.
\ee

As a complex manifold ($\mathbb{P}^1\times\mathbb{P}^1$) the metric
is written as two copies of the Fubini-Study metric
\be\label{eq:complexmetric}
\begin{aligned}
ds^2&=2g_{z\bar z}(z,\bar z)dz d\bar z+2g_{w\bar w}(w, \bar w)dwd\bar w \\
    &=4r_1^2\frac{dzd\bar z}{(1+|z|^2)^2}+4r_2^2\frac{dwd\bar w}{(1+|w|^2)^2},
\end{aligned}
\ee
The change of variables from real to complex coordinates is
\be
z=\tan(\theta_1/2)e^{i\varphi_1}, \qquad w=\tan(\theta_2/2)e^{i\varphi_2}.
\ee

The flat metric in complex coordinate has the following nonzero components
$\delta_{1\bar 1}=\delta_{2\bar 2}=\frac{1}{2}$, $\delta^{1\bar 1}=\delta^{2\bar 2}=2$.
Then defining $\sqrt{g_1}:=2g_{z\bar z}$ and $\sqrt{g_2}:=2g_{w\bar w}$
we can write rewrite the metric (\ref{eq:complexmetric}) using complex vierbein 1-forms
\be
ds^2=e^1 e^{\bar 1} + e^2 e^{\bar 2}
\ee
where
\be
e^1=g_1^{1/4}dz, \quad e^{\bar 1}=g_1^{1/4}d\bar z, \quad e^2=g_2^{1/4}dw, \quad e^{\bar 2}=g_2^{1/4}d\bar w.
\ee
Moreover we can write the non-zero Christoffel symbols of the Levi-Civita connection as
\be
\begin{aligned}
&\Gamma_{zz}^z=\frac{1}{2}\de_z \log g_1, \quad\;\; \Gamma_{\bar z \bar z}^{\bar z}=\frac{1}{2}\de_{\bar z} \log g_1, \\
&\Gamma_{ww}^w=\frac{1}{2}\de_w \log g_2, \quad \Gamma_{\bar w \bar w}^{\bar w}=\frac{1}{2}\de_{\bar w} \log g_2,
\end{aligned}
\ee
and the relative spin connection\footnote{
The convention for the spin connection is
$\omega_\m^{ab}=e^{[a}_\n\de_\m e^{b]\n}+e^{[a}_\n e^{b]\rho}\Gamma_{\m\rho}^\n$.}
$\omega_\m:=-2i\omega_{\m 1\bar1}$,
$\omega'_\m:=-2i\omega_{\m 2\bar2}$ as
\be\label{Spin-Conn}
\begin{aligned}
&\omega_z=\frac{i}{4}\de_z \log g_1, \quad\;\; \omega_{\bar z}=-\frac{i}{4}\de_{\bar z} \log g_1, \\
&\omega'_w=\frac{i}{4}\de_w \log g_2, \quad \omega'_{\bar w}=-\frac{i}{4}\de_{\bar w} \log g_2.
\end{aligned}
\ee
The explicit expression for the non zero components of the spin connection in complex coordinates are
\be
\omega_z=-i\frac{\bar z}{1+|z|^2}, \quad \omega_{\bar z}=i\frac{z}{1+|z|^2}, \quad
\omega'_w=-i\frac{\bar w}{1+|w|^2}, \quad \omega'_{\bar w}=i\frac{w}{1+|w|^2}.
\ee
If one prefers to work in polar coordinates, the spin connection
($\omega_\m=\omega_{\m 12}$, $\omega'_\m=\omega_{\m 34}$) is
\be
\omega_{\theta_1}=\omega'_{\theta_2}=0, \quad \omega_{\varphi_1}=-\cos\theta_1, \quad \omega'_{\varphi_2}=-\cos\theta_2.
\ee
The Riemann tensor $R_{\m\n\rho\s}$ has two independent components
\be
R_{z\bar z z\bar z}=-\frac{1}{2}g_{z\bar z}g_{z\bar z} \mathcal{R}_1, \qquad
R_{w\bar w w\bar w}=-\frac{1}{2}g_{w\bar w}g_{w\bar w} \mathcal{R}_2,
\ee
where $\mathcal{R}_1$ and $\mathcal{R}_2$ are the summands of the Riemann scalar
$\mathcal{R}=\mathcal{R}_1+\mathcal{R}_2$
related respectively to the first and second sphere, which are expressed as
\be
\mathcal{R}_1=-\frac{2}{\sqrt{g_1}}\de_z \de_{\bar z}\log g_1=\frac{2}{r_1^2}, \qquad
\mathcal{R}_2=-\frac{2}{\sqrt{g_2}}\de_w \de_{\bar w}\log g_2=\frac{2}{r_2^2}.
\ee
Finally using the spin connection is possible to write the action of the covariant derivative
on spinors
\be\label{eq:Dspinor}
\nabla_\m \psi=\big(\de_\m+\frac{i}{2}(\omega_\m-\omega'_\mu)\sigma_3\big)\psi,
\qquad
\nabla_\m\chir\psi=\big(\de_\m+\frac{i}{2}(\omega_\m+\omega'_\mu)\sigma_3\big)\chir\psi.
\ee
And on 1-forms
\be\label{eq:D1form}
\nabla_\m X_a=e_a^\n \nabla_\m X_\n
\ee
where $\nabla_\m$ on the r.h.s. is the Levi-Civita connection.

\subsection{Spinor convention}

Left and right chirality spinors are denoted $\xi_{A\alpha}$ and $\chir\xi_{A}^{\phantom{A}\dot\alpha}$.
The multiplication of spinors is usually implicit as $\xi^A\xi_A=\xi^{A\alpha}\xi_{A\alpha}=\ve^{AB}\ve^{\alpha\beta}\xi_{B\beta}\xi_{A\alpha}$
and $\chir\xi_A\chir\xi^A=\chir\xi_{A\dot\alpha}\chir\xi^{A\dot\alpha}=\ve^{AB}\ve_{\dot\alpha\dot\beta}\chir\xi_{A}^{\phantom{A}\dot\beta}\chir\xi_{B}^{\phantom{B}\dot\alpha}$.
The invariant antisymmetric tensors are $\ve^{\alpha\beta}$, $\ve_{\alpha\beta}$ for left chirality spinors,
$\ve^{\dot\alpha\dot\beta}$, $\ve_{\dot\alpha\dot\beta}$ for right chirality ones,
with $\ve^{12}=1$ $\ve_{12}=-1$.
Our choice for the two set of matrices $(\s_a)_{\alpha\dot\alpha}$, $(\chir\s_a)^{\dot\alpha\alpha}$
($a=1,2,3,4$) is
\be\label{eq:sigmaflat}
\begin{aligned}
\s_a      &=\{-i\s_j,\Id\}, \quad j=1,2,3, \\
\chir\s_a &=\{+i\s_j,\Id\}, \quad j=1,2,3,
\end{aligned}
\ee
where $\s_j$ are the Pauli matrices.
Their expression with curved indices is derived using vierbein:
\be
\s_\m=e_\m^a \s_a, \qquad \chir\s_\m=e_\m^a \chir\s_a.
\ee

The generators of rotations for left and right chirality spinors are respectively
\be
\s_{ab}=\frac{1}{2}(\s_a\chir\s_b-\s_b\chir\s_a), \qquad
\chir\s_{ab}=\frac{1}{2}(\chir\s_a\s_b-\chir\s_b\s_a).
\ee
Note that $\s_{ab}$ is anti self-dual and $\chir\s_{ab}$ is self-dual.

Some useful identities of the sigma matrices are
\be\label{eq:sigmaIds1}
\begin{aligned}
&\s_a\chir\s_b+\s_b\chir\s_a=2\delta_{ab}, \\
&\chir\s_a\s_b+\chir\s_b\s_a=2\delta_{ab}, \\
&\s_a\chir\s_b\s_c=\delta_{ab}\s_c+\delta_{bc}\s_a-\delta_{ac}\s_b+\ve_{abcd}\s^d, \\
&\chir\s_a \s_b \chir\s_c=\delta_{ab}\chir\s_c+\delta_{bc}\chir\s_a-\delta_{ac}\chir\s_b-\ve_{abcd}\chir\s^d.
\end{aligned}
\ee
The last two identities imply
\be\label{eq:sigmaIds2}
\s_{ab}\s_c=-4P_{abcd}^-\s^d, \qquad \chir\s_{ab}\chir\s_c=-4P_{abcd}^+ \chir\s^d,
\ee
where $P^-$ and $P^+$ are given by
\be\label{eq:Projpm}
P_{abcd}^\pm=\frac{1}{4}(\delta_{ac}\delta_{bd}-\delta_{ad}\delta_{bc}\pm\ve_{abcd}).
\ee
and are projectors on the anti-self dual and self dual forms respectively.

Others useful identities are
\be\label{eq:sigmaIds2bis}
\begin{aligned}
&(\chir\s_\m)^{\dot\alpha\alpha}(\s^\m)_{\beta\dot\beta}=2\delta^{\alpha}_{\beta}\delta^{\dot\alpha}_{\dot\beta}, \\
&(\s_\m)_{\alpha\dot\alpha}(\s^\m)_{\beta\dot\beta}=2\ve_{\alpha\beta}\ve_{\dot\alpha\dot\beta}, \\
&(\chir\s_\m)^{\dot\alpha\alpha}(\chir\s^\m)^{\dot\beta\beta}=2\ve^{\alpha\beta}\ve^{\dot\alpha\dot\beta}.
\end{aligned}
\ee

\section{Special functions}
\label{App:special-fun}

The Barnes' double zeta function $\zeta_2$ has the following integral representation:
\be\label{zeta2}
\zeta_2(x;s|\ve_1,\ve_2)=\frac{1}{\Gamma(s)}\int_0^\infty dt\,t^{s-1}\frac{e^{-tx}}{(1-e^{-\ve_1 t})(1-e^{-\ve_2 t})}.
\ee
This integral is well-defined if $\text{Re}\,\ve_1>0$, $\text{Re}\,\ve_2>0$, $\text{Re}\,x>0$
and can be analytically continued to all complex values of $\ve_1$ and $\ve_2$
except when $\frac{\ve_1}{\ve_2}\neq a$ with $a\in\mathbb{R}_{<0}$.
The following series expansion of $\zeta_2$ for $\ve_1>0$, $\ve_2>0$ is convergent if $\text{Re}\,s>2$:
\be\label{serieszeta2}
\zeta_2(x;s|\ve_1,\ve_2)=\sum_{m,n\ge 0}(x+m\ve_1+n\ve_2)^{-s}.
\ee
The Barnes' double Gamma function $\Gamma_2$, defined by
\be\label{Gamma2}
\log\Gamma_2(x|\ve_1,\ve_2)=\frac{d}{d s}\Big|_{s=0}\zeta_2(x;s|\ve_1,\ve_2),
\ee
is analytic in $x$ except at the poles at $x=-m\ve_1 -n\ve_2 $ with $m,n\in\mathbb{Z}$.
Define
\be\label{gamma2}
\gamma_{\ve_1,\ve_2}(x)=\frac{d}{d s}\Big|_{s=0}\frac{1}{\Gamma(s)}
                     \int_0^\infty dt\,t^{s-1}\frac{e^{-tx}}{(1-e^{\ve_1 t})(1-e^{\ve_2 t})}.
\ee
Then
\be\label{RelGamma2gamma2}
\gamma_{\ve_1,\ve_2}(x)=\log\Gamma_2(x|-\ve_1,-\ve_2).
\ee
The function $\Gamma_2$ has the following infinite-product representations:
\be\label{prodGamma2}
\Gamma_2(x|\ve_1,\ve_2)=\left\{
\begin{aligned}
&\prod_{m,n\ge 0}\big(x+m\ve_1+n\ve_2\big)^{-1} &\qquad& \text{if $\ve_1>0$, $\ve_2>0$,} \\
&\prod_{m,n\ge 0}\big(x+m\ve_1-(n-1)\ve_2\big) &\qquad& \text{if $\ve_1>0$, $\ve_2<0$,} \\
&\prod_{m,n\ge 0}\big(x-(m-1)\ve_1+n\ve_2\big) &\qquad& \text{if $\ve_1<0$, $\ve_2>0$,} \\
&\prod_{m,n\ge 0}\big(x-(m-1)\ve_1-(n-1)\ve_2\big)^{-1} &\qquad& \text{if $\ve_1<0$, $\ve_2<0$.}
\end{aligned}
\right.
\ee
The function $\Gamma_2$ satisfies the following multiplicative identity
\be
\Gamma_2(x+\ve_1|\ve_1,\ve_2)\Gamma_2(x+\ve_2|\ve_1,\ve_2)
	=x\Gamma_2(x|\ve_1,\ve_2)\Gamma_2(x+Q|\ve_1,\ve_2)
\ee
where $Q=\ve_1+\ve_2$, the shift identities
\be\label{Gamma2shift}
\begin{aligned}
&\Gamma_2(x+\ve_1|\ve_1,\ve_2)=\frac{\sqrt{2\pi}\ve_2^{1/2-x/\ve_2}}{\Gamma(x/\ve_2)}\Gamma_2(x|\ve_1,\ve_2), \\
&\Gamma_2(x+\ve_2|\ve_1,\ve_2)=\frac{\sqrt{2\pi}\ve_1^{1/2-x/\ve_1}}{\Gamma(x/\ve_1)}\Gamma_2(x|\ve_1,\ve_2).
\end{aligned}
\ee
The Upsilon function is defined as
\be\label{Upsilon}
\Upsilon_{\ve_1,\ve_2}(x)=\frac{1}{\Gamma_2(x|\ve_1,\ve_2)\Gamma_2(Q-x|\ve_1,\ve_2)}=\Upsilon_{\ve_1,\ve_2}(Q-x)
\ee
It exhibits the shift property
\be\label{Upsilonshift}
\begin{aligned}
&\Upsilon_{\ve_1,\ve_2}(x+\ve_1)=\ve_2^{2x/\ve_2-1}\gamma(x/\ve_2)\Upsilon_{\ve_1,\ve_2}(x), \\
&\Upsilon_{\ve_1,\ve_2}(x+\ve_2)=\ve_1^{2x/\ve_1-1}\gamma(x/\ve_1)\Upsilon_{\ve_1,\ve_2}(x),
\end{aligned}
\ee
where
\be\label{gamma}
\gamma(x)=\frac{\Gamma(x)}{\Gamma(1-x)}.
\ee


\bibliography{REf}

\providecommand{\href}[2]{#2}\begingroup\raggedright\begin{thebibliography}{10}

\bibitem{Dumitrescu:2012ha}
T.~T. Dumitrescu, G.~Festuccia, and N.~Seiberg, {\it {Exploring Curved
  Superspace}},  {\em JHEP} {\bf 1208} (2012) 141,
  [\href{http://arxiv.org/abs/1205.1115}{{\tt arXiv:1205.1115}}].

\bibitem{Dumitrescu:2012at}
T.~T. Dumitrescu and G.~Festuccia, {\it {Exploring Curved Superspace (II)}},
  {\em JHEP} {\bf 1301} (2013) 072, [\href{http://arxiv.org/abs/1209.5408}{{\tt
  arXiv:1209.5408}}].

\bibitem{Hama:2012bg}
N.~Hama and K.~Hosomichi, {\it {Seiberg-Witten Theories on Ellipsoids}},  {\em
  JHEP} {\bf 1209} (2012) 033, [\href{http://arxiv.org/abs/1206.6359}{{\tt
  arXiv:1206.6359}}].

\bibitem{Klare:2013dka}
C.~Klare and A.~Zaffaroni, {\it {Extended Supersymmetry on Curved Spaces}},
  {\em JHEP} {\bf 1310} (2013) 218, [\href{http://arxiv.org/abs/1308.1102}{{\tt
  arXiv:1308.1102}}].

\bibitem{Gupta:2012cy}
R.~K. Gupta and S.~Murthy, {\it {All solutions of the localization equations
  for N=2 quantum black hole entropy}},  {\em JHEP} {\bf 1302} (2013) 141,
  [\href{http://arxiv.org/abs/1208.6221}{{\tt arXiv:1208.6221}}].

\bibitem{Witten:1988ze}
E.~Witten, {\it {Topological Quantum Field Theory}},  {\em Commun.Math.Phys.}
  {\bf 117} (1988) 353.

\bibitem{Witten:1991zz}
E.~Witten, {\it {Mirror manifolds and topological field theory}},
  \href{http://arxiv.org/abs/hep-th/9112056}{{\tt hep-th/9112056}}.

\bibitem{Nekrasov:2002qd}
N.~A. Nekrasov, {\it {Seiberg-Witten prepotential from instanton counting}},
  {\em Adv.Theor.Math.Phys.} {\bf 7} (2004) 831--864,
  [\href{http://arxiv.org/abs/hep-th/0206161}{{\tt hep-th/0206161}}].

\bibitem{Bruzzo:2002xf}
U.~Bruzzo, F.~Fucito, J.~F. Morales, and A.~Tanzini, {\it {Multiinstanton
  calculus and equivariant cohomology}},  {\em JHEP} {\bf 0305} (2003) 054,
  [\href{http://arxiv.org/abs/hep-th/0211108}{{\tt hep-th/0211108}}].

\bibitem{Nekrasov:2003vi}
N.~Nekrasov, ``{Localizing gauge theories}.''
  \url{http://www.researchgate.net/publication/253129819_ Localizing_ gauge_
  theories}, 2003.

\bibitem{Bonelli:2012ny}
G.~Bonelli, K.~Maruyoshi, A.~Tanzini, and F.~Yagi, {\it {N=2 gauge theories on
  toric singularities, blow-up formulae and W-algebrae}},  {\em JHEP} {\bf
  1301} (2013) 014, [\href{http://arxiv.org/abs/1208.0790}{{\tt
  arXiv:1208.0790}}].

\bibitem{Alday:2009aq}
L.~F. Alday, D.~Gaiotto, and Y.~Tachikawa, {\it {Liouville Correlation
  Functions from Four-dimensional Gauge Theories}},  {\em Lett.Math.Phys.} {\bf
  91} (2010) 167--197, [\href{http://arxiv.org/abs/0906.3219}{{\tt
  arXiv:0906.3219}}].

\bibitem{Bruzzo:2013daa}
U.~Bruzzo, M.~Pedrini, F.~Sala, and R.~J. Szabo, {\it {Framed sheaves on root
  stacks and supersymmetric gauge theories on ALE spaces}},
  \href{http://arxiv.org/abs/1312.5554}{{\tt arXiv:1312.5554}}.

\bibitem{Bruzzo:2014jza}
U.~Bruzzo, F.~Sala, and R.~J. Szabo, {\it {N=2 quiver gauge theories on A-type
  ALE spaces}},  \href{http://arxiv.org/abs/1410.2742}{{\tt arXiv:1410.2742}}.

\bibitem{Samtleben:2012gy}
H.~Samtleben and D.~Tsimpis, {\it {Rigid supersymmetric theories in 4d
  Riemannian space}},  {\em JHEP} {\bf 1205} (2012) 132,
  [\href{http://arxiv.org/abs/1203.3420}{{\tt arXiv:1203.3420}}].

\bibitem{Razamat:2013jxa}
S.~S. Razamat and M.~Yamazaki, {\it {S-duality and the N=2 Lens Space Index}},
  {\em JHEP} {\bf 1310} (2013) 048, [\href{http://arxiv.org/abs/1306.1543}{{\tt
  arXiv:1306.1543}}].

\bibitem{Bobev:2013cja}
N.~Bobev, H.~Elvang, D.~Z. Freedman, and S.~S. Pufu, {\it {Holography for $N =
  2^*$ on $S^4$}},  {\em JHEP} {\bf 1407} (2014) 001,
  [\href{http://arxiv.org/abs/1311.1508}{{\tt arXiv:1311.1508}}].

\bibitem{Closset:2013sxa}
C.~Closset and I.~Shamir, {\it {The $\mathcal{N}=1$ Chiral Multiplet on
  $T^2\times S^2$ and Supersymmetric Localization}},  {\em JHEP} {\bf 1403}
  (2014) 040, [\href{http://arxiv.org/abs/1311.2430}{{\tt arXiv:1311.2430}}].

\bibitem{Assel:2014paa}
B.~Assel, D.~Cassani, and D.~Martelli, {\it {Localization on Hopf surfaces}},
  {\em JHEP} {\bf 1408} (2014) 123, [\href{http://arxiv.org/abs/1405.5144}{{\tt
  arXiv:1405.5144}}].

\bibitem{Gomis:2014woa}
J.~Gomis and N.~Ishtiaque, {\it {Kahler Potential and Ambiguities in 4d N=2
  SCFTs}},  \href{http://arxiv.org/abs/1409.5325}{{\tt arXiv:1409.5325}}.

\bibitem{Marmiroli:2014ssa}
D.~Marmiroli, {\it {Phase structure of $\mathcal{N}=2^*$ SYM on ellipsoids}},
  \href{http://arxiv.org/abs/1410.4715}{{\tt arXiv:1410.4715}}.

\bibitem{GM}
J.~Gomis and C.~Musema, {\it {Partition function of N=2 Gauge Theories on
  $S^2\times S^2$}},  {\em in preparation}.

\bibitem{Bonelli:2009zp}
G.~Bonelli and A.~Tanzini, {\it {Hitchin systems, $N=2$ gauge theories and
  W-gravity}},  {\em Phys.Lett.} {\bf B691} (2010) 111--115,
  [\href{http://arxiv.org/abs/0909.4031}{{\tt arXiv:0909.4031}}].

\bibitem{Alday:2009qq}
L.~F. Alday, F.~Benini, and Y.~Tachikawa, {\it {Liouville/Toda central charges
  from M5-branes}},  {\em Phys.Rev.Lett.} {\bf 105} (2010) 141601,
  [\href{http://arxiv.org/abs/0909.4776}{{\tt arXiv:0909.4776}}].

\bibitem{Klare:2012gn}
C.~Klare, A.~Tomasiello, and A.~Zaffaroni, {\it {Supersymmetry on Curved Spaces
  and Holography}},  {\em JHEP} {\bf 1208} (2012) 061,
  [\href{http://arxiv.org/abs/1205.1062}{{\tt arXiv:1205.1062}}].

\bibitem{Closset:2014pda}
C.~Closset and S.~Cremonesi, {\it {Comments on $ \mathcal{N} = (2, 2)$
  supersymmetry on two-manifolds}},  {\em JHEP} {\bf 1407} (2014) 075,
  [\href{http://arxiv.org/abs/1404.2636}{{\tt arXiv:1404.2636}}].

\bibitem{Baulieu:2005bs}
L.~Baulieu, G.~Bossard, and A.~Tanzini, {\it {Topological vector symmetry of
  BRSTQFT and construction of maximal supersymmetry}},  {\em JHEP} {\bf 0508}
  (2005) 037, [\href{http://arxiv.org/abs/hep-th/0504224}{{\tt
  hep-th/0504224}}].

\bibitem{Bershtein:2013oka}
M.~Bershtein, B.~Feigin, and A.~Litvinov, {\it {Coupling of two conformal field
  theories and Nakajima-Yoshioka blow-up equations}},
  \href{http://arxiv.org/abs/1310.7281}{{\tt arXiv:1310.7281}}.

\bibitem{Zamolodchikov:2005fy}
A.~B. Zamolodchikov, {\it {On the three-point function in minimal Liouville
  gravity}},  \href{http://arxiv.org/abs/hep-th/0505063}{{\tt hep-th/0505063}}.

\bibitem{0510214}
A.~Belavin and A.~Zamolodchikov, {\it {Polyakov's string: Twenty five years
  after. Proceedings}},  \href{http://arxiv.org/abs/hep-th/0510214}{{\tt
  hep-th/0510214}}.

\bibitem{RS}
S.~Ribault and R.~Santachiara, {\it Liouville theory with a central charge less
  than one},  {\em to appear}.

\bibitem{Bonelli:2011jx}
G.~Bonelli, K.~Maruyoshi, and A.~Tanzini, {\it {Instantons on ALE spaces and
  Super Liouville Conformal Field Theories}},  {\em JHEP} {\bf 1108} (2011)
  056, [\href{http://arxiv.org/abs/1106.2505}{{\tt arXiv:1106.2505}}].

\bibitem{Bonelli:2011kv}
G.~Bonelli, K.~Maruyoshi, and A.~Tanzini, {\it {Gauge Theories on ALE Space and
  Super Liouville Correlation Functions}},  {\em Lett.Math.Phys.} {\bf 101}
  (2012) 103--124, [\href{http://arxiv.org/abs/1107.4609}{{\tt
  arXiv:1107.4609}}].

\bibitem{Belavin:2011sw}
A.~Belavin, M.~Bershtein, B.~Feigin, A.~Litvinov, and G.~Tarnopolsky, {\it
  {Instanton moduli spaces and bases in coset conformal field theory}},  {\em
  Comm. Math. Phys. 319 1, pp} {\bf 269-301} (2013) 269--301,
  [\href{http://arxiv.org/abs/1111.2803}{{\tt arXiv:1111.2803}}].

\bibitem{Schomerus:2012se}
V.~Schomerus and P.~Suchanek, {\it {Liouville's Imaginary Shadow}},
  \href{http://arxiv.org/abs/1210.1856}{{\tt arXiv:1210.1856}}.

\bibitem{Hadasz:2013dza}
L.~Hadasz and Z.~Jaskólski, {\it {Super-Liouville - Double Liouville
  correspondence}},  {\em JHEP} {\bf 1405} (2014) 124,
  [\href{http://arxiv.org/abs/1312.4520}{{\tt arXiv:1312.4520}}].

\bibitem{Benini:2012ui}
F.~Benini and S.~Cremonesi, {\it {Partition functions of $\mathcal{N}=(2,2)$
  gauge theories on $S^2$ and vortices}},
  \href{http://arxiv.org/abs/1206.2356}{{\tt arXiv:1206.2356}}.

\bibitem{Doroud:2012xw}
N.~Doroud, J.~Gomis, B.~Le~Floch, and S.~Lee, {\it {Exact Results in $D=2$
  Supersymmetric Gauge Theories}},  {\em JHEP} {\bf 1305} (2013) 093,
  [\href{http://arxiv.org/abs/1206.2606}{{\tt arXiv:1206.2606}}].

\bibitem{Alday:2009fs}
L.~F. Alday, D.~Gaiotto, S.~Gukov, Y.~Tachikawa, and H.~Verlinde, {\it {Loop
  and surface operators in N=2 gauge theory and Liouville modular geometry}},
  {\em JHEP} {\bf 1001} (2010) 113, [\href{http://arxiv.org/abs/0909.0945}{{\tt
  arXiv:0909.0945}}].

\bibitem{Drukker:2009id}
N.~Drukker, J.~Gomis, T.~Okuda, and J.~Teschner, {\it {Gauge Theory Loop
  Operators and Liouville Theory}},  {\em JHEP} {\bf 1002} (2010) 057,
  [\href{http://arxiv.org/abs/0909.1105}{{\tt arXiv:0909.1105}}].

\bibitem{Awata:2010bz}
H.~Awata, H.~Fuji, H.~Kanno, M.~Manabe, and Y.~Yamada, {\it {Localization with
  a Surface Operator, Irregular Conformal Blocks and Open Topological String}},
   {\em Adv.Theor.Math.Phys.} {\bf 16} (2012), no.~3 725--804,
  [\href{http://arxiv.org/abs/1008.0574}{{\tt arXiv:1008.0574}}].

\end{thebibliography}\endgroup
\bibliographystyle{JHEP}

  
\end{document}